\documentclass[galaxies,article,accept,pdftex,moreauthors]{Definitions/mdpi} 

\firstpage{1} 
\pubvolume{1}
\issuenum{1}
\articlenumber{0}
\pubyear{2026}
\copyrightyear{2026}
\externaleditor{Firstname Lastname} 
\datereceived{2 June 2026 } 
\daterevised{30 July 2026 } 
\dateaccepted{5 August 2026 } 
\datepublished{ } 

\usepackage{bm}
\usepackage{xcolor}
\usepackage{longtable}
\newcommand\ion[2]{\textup{#1\,\textsc{\lowercase{#2}}}} 
  
\newcommand{\apjs}{Astrophys. J. Suppl. Ser. }
\newcommand{\apjl}{Astrophys. J. Lett. }
\newcommand{\mnras}{Mon. Not. R. Astron. Soc.}

\newcommand{\pasp}{Publ. Astron. Soc. Pac. }
\newcommand{\pasa}{Publ. Astron. Soc. Aust. }

\newcommand{\araa}{Annu. Rev. Astron. Astrophys. }

\Title{Radio Properties of Narrow-Line and Broad-Line Seyfert 1~Galaxies}

\Author{Krisztina \'E. Gab\'anyi $^{1,2,3,4,}$*\orcidA{}, S. Komossa $^{5}$\orcidB{}, Attila Mez\H{o}si $^{1}$\orcidC{}, Emese Forg\'acs-Dajka $^{1,2,6}$\orcidD{} and S\'andor Frey $^{1,2}$ \orcidE{}}

\AuthorNames{K. \'E. Gab\'anyi, S. Komossa, A. Mez\H{o}si, E. Forg\'acs-Dajka and S. Frey}

\address{%
$^{1}$ \quad Department of  Astronomy, Institute of Physics and Astronomy, ELTE E\"otv\"os Lor\'and University,\linebreak  
P\'azm\'any P\'eter s\'et\'any 1/A, 1117 Budapest, Hungary; ytserrof@student.elte.hu (A.M.); forgacs.emese@ttk.elte.hu (E.F.-D.); frey.sandor@csfk.org (S.F.) \\
$^{2}$ \quad Konkoly Observatory, HUN-REN Research Centre for Astronomy and Earth Sciences, MTA Centre of Excellence, Konkoly-Thege Mikl\'os \'ut 15-17, 1121 Budapest, Hungary \\
$^{3}$ \quad HUN-REN--ELTE Extragalactic Astrophysics Research Group, ELTE E\"otv\"os Lor\'and University,\linebreak  
P\'azm\'any P\'eter s\'et\'any 1/A, 1117 Budapest, Hungary \\
$^{4}$ \quad Institute of Astronomy, Faculty of Physics, Astronomy and Informatics, Nicolaus Copernicus University, Grudzi\c{a}dzka 5, 87-100 Toru\'n, Poland \\
$^{5}$ \quad Max-Planck-Institut f\"ur Radioastronomie, Auf dem H\"ugel 69, D-53121 Bonn, Germany; skomossa@mpifr.de\\
$^{6}$ \quad HUN-REN--SZTE Stellar Astrophysics Research Group, 766 Szegedi Rd., 6500 Baja, Hungary}

\corres{Correspondence: k.gabanyi@astro.elte.hu}

\abstract{Narrow-line Seyfert 1 (NLS1) galaxies host active galactic nuclei (AGN) with narrow optical emission lines of the broad-line region. This is often explained with a relatively lower mass of the central supermassive black hole and super-Eddington accretion. We compared the radio properties of large samples of NLS1 and broad-line Seyfert 1 (BLS1) galaxies compiled from the Sloan Digital Sky Survey. We cross-matched the NLS1 and BLS1 samples with the Faint Images of the Radio Sky at Twenty-Centimeters (FIRST) sky survey at $1.4$\,GHz and the first and second epoch data of the Very Large Array Sky Survey (VLASS) at $3$\,GHz. We calculated the radio spectral indices, the $1.4$-GHz radio power, and the radio loudness. We found lower $1.4$-GHz radio detection rates for the NLS1 galaxies. The median radio loudness values, the fraction of radio-loud AGN, and the median $1.4$-GHz radio power are also lower for the NLS1 sample. The median spectral indices imply a slightly steeper radio spectrum for the NLS1 sample than for the BLS1 sample. Comparison of the star formation rates estimated from the radio data and the infrared measurements of the Wide-field Infrared Survey Explorer satellite indicated that more than half of the FIRST- and VLASS-detected NLS1 and BLS1 galaxies contain radio-emitting AGN.}


\begin{document}


\section{Introduction}

Narrow-line Seyfert 1 galaxies (NLS1) are a peculiar subgroup of active galactic nuclei (AGN) showing narrow optical emission lines from the broad-line region (BLR). The~full-width at half maximum (FWHM) of the broad H$\beta$ line is $<$2000\,km\,s$^{-1}$, the~flux ratio of the [\ion{O}{III}]$\lambda$5007 to H$\beta$ line is $<$3 \cite{Osterbrock1985, Goodrich1989}. They are often characterized by strong emission of the \ion{Fe}{II} multiplets~\cite{Veron2004}. Such properties are explained by the high accretion rate of a relatively smaller mass supermassive black hole (see a review by, e.g., \cite{Komossa2008}). 

While most NLS1 galaxies are radio-quiet, their radio properties are diverse, ranging from weak radio emitters to rare systems hosting powerful relativistic jets. NLS1s are more radio-quiet than the general AGN population, with~$\sim$7\% of them being radio-loud~\cite{Komossa2006, varglund2025}. At~low radio frequencies, a~recent study~\cite{varglund2025} reported detection rates as high as $40\%$ at $144$\,MHz in the Low-Frequency Array (LOFAR) Two-metre Sky Survey (LoTSS,  \cite{lofar_dr2}) Data Release 3 (DR3). Additionally, there are a number of extremely radio-loud NLS1 galaxies showing properties similar to blazars, AGN with jets pointing close to the line of sight, such as relativistic jets, and even gamma-ray emission (e.g., ~Refs. \cite{Yuan2008, Komossa2006, Foschni2022, multi_gk_2025} and references therein). 

The interpretation of the radio properties of NLS1 galaxies can be complicated by selection effects and by ambiguities in the optical classification of large samples~\cite{varglund2025,Marziani2018rnls.confE...2M}. Therefore, a~comparison with an appropriate control sample is essential when assessing whether the observed trends reflect intrinsic differences between narrow- and broad-line type 1 AGN or arise from sample selection. This is particularly important when radio loudness and related quantities are investigated, since extinction and reddening can affect the optical measurements and thereby bias their interpretation~\cite{Komossa2006,paliya2024}.

Recently, Ref.~\cite{paliya2024} published catalogs of NLS1 and broad-line Seyfert 1 galaxies (BLS1) based on the spectra of the Sloan Digital Sky Survey Data Release 17 (SDSS DR17, ~\cite{sdssdr17}). They considered objects classified as either galaxies or quasars with redshift $z<0.8$ to make sure that both the H$\beta$ and the [\ion{O}{III}] emission lines fall within the spectral range of SDSS and excluded objects with low signal-to-noise ratio (SNR) or~with otherwise flagged data. They updated the SDSS pipeline-based redshift values with the improved redshift measurements of the SDSS DR16~\cite{wushen}. 
They used the publicly available Bayesian AGN Decomposition Analysis for SDSS Spectra (BADASS,~\cite{badass}) code to classify these sources. Most of the objects were type 2 AGN; the~remainder, nearly $75{,}000$ objects, were classified into either NLS1 or BLS1 categories. This classification was based on the FWHM of the broad H$\beta$ emission line and the flux ratio of the [\ion{O}{III}] and H$\beta$ emission lines. Thus, their final NLS1 catalog contained $22{,}656$ objects, while the other $52{,}273$ were classified as BLS1~galaxies.

The radio loudness of the samples was investigated by cross-matching them with the Faint Images of the Radio Sky at Twenty-Centimeters (FIRST,~\cite{first_white}) survey data and assuming the same spectral index for all objects~\cite{paliya2024}. 
Here, we extend this comparison by investigating the radio properties of the NLS1 and BLS1 samples in a uniform framework using the FIRST as well as the more recent Very Large Array Sky Survey (VLASS,~\cite{vlass_lacy}). Our aim is to test whether the two classes differ in their radio detection rate, spectral index, radio loudness, and~$1.4$-GHz radio power. We also assess to what extent the observed radio emission can be related to AGN activity rather than star~formation.

Paliya~et~al. (2024) \cite{paliya2024} noted that they did not discriminate their objects by optical magnitude. However, they list the absolute $B$-band magnitude ($M_B$) in their catalogs. Therefore, when comparing their data in the BLS1 and NLS1 catalogs, we divided them into quasar and Seyfert galaxy populations using the limiting value of $M_B=-23$ \cite{SchmidtGreen1983}. In the following, we will refer to the two catalogs of~\cite{paliya2024} as BL1 and NL1 containing the broad-line type 1 and the narrow-line type 1 AGN, respectively. We will differentiate between the optically brighter and fainter subsamples as type 1 quasars and Seyfert 1 galaxies, respectively. We will refer to these subsamples as narrow-line type 1 quasars or NLQ1, narrow-line Seyfert 1 galaxies or NLS1 galaxies, broad-line type 1 quasars or BLQ1, and~broad-line Seyfert 1 galaxies or BLS1 galaxies.

The paper is structured as follows. In~Section~\ref{sec:samples}, we describe the samples, the~radio cross-matching procedure, and~the construction of the subsamples used in the analysis. In~Section~\ref{sec:results}, we present the main radio properties of the NL1 and BL1 sources, including their detection rates, spectral indices, radio loudness, radio power, and~compactness. In~Section~\ref{sec:discussion}, we discuss the implications of these results, with~particular emphasis on the origin of the radio emission and on the role of AGN activity and star formation. In~Section~\ref{sec:summary}, we summarize our main conclusions. During~the analysis, we identified a few individual sources with extremely large spectral indices or radio loudness. These are listed in Appendices~\ref{sec:alpha_outlier} and \ref{sec:extremeRL}, respectively. In~the following, we assume a flat $\Lambda$CDM cosmological model with $H_0=70$\,km\,s$^{-1}$\,Mpc$^{-1}$ and~$\Omega_\mathrm{m}=0.3$. We use the radio spectral index $\alpha$ defined as $S \propto \nu^\alpha$, where $S$ is the flux density, and $\nu$ is the observing~frequency.

\section{The~Samples} \label{sec:samples}

In the BL1 and NL1 catalogs of~\cite{paliya2024}, there are $52{,}273$ and $22{,}656$ objects, respectively. However, we found that there are $276$ duplicate entries that refer to the same optical objects in the BL1 catalog. Thus, there are  $51{,}997$ unique objects in the BL1 catalog.
Similarly, we found $63$ duplicates in the NL1 catalog. Later, we found three objects among the five extremely radio-loud ones in the NL1 sample whose optical spectra could be better described with heavily-obscured AGN; therefore, we removed them from the following analysis (for more details, see Appendix~\ref{sec:extremeRL}). Thus, $22{,}590$ unique NL1 AGN remained.

Following~\cite{paliya2024}, we cross-matched their catalogs with the FIRST using a search radius of $5^{\prime\prime}$. We also checked within the FIRST catalog whether there are multiple detections within this search radius. We found $13$ and $1$ such objects in the BL1 and NL1 samples, respectively. We inspected the FIRST cutout images of the objects having multiple FIRST counterparts and~found that in all but two cases (both BL1 AGN), the~multiple radio emissions can most probably be connected to a single optical galaxy, the~one listed in the catalog of~\cite{paliya2024}. For~those objects, we included the summed flux density of the FIRST counterparts in our final~catalog.

The two exceptions are SDSS\,J161318.18$+$341544.0 and SDSS\,J1222517.95$+$222018.6. The~first one showed a rather diffuse radio structure in its FIRST image described by two components in the catalog. However, both of those have rather high sidelobe probability, $0.526$ and $0.282$; therefore, we did not include this object as a radio detection. For~SDSS\,J1222517.95$+$222018.6, we retained as the FIRST counterpart the detection that has lower sidelobe probability ($0.016$ versus $0.027$). Thus, we recovered $727$ and $2567$ FIRST counterparts for the NL1 and the BL1 samples, respectively. 

Principally, there could be chance coincidences among these counterparts. To~estimate the probability of finding unrelated radio emissions to the optical galaxies within the FIRST catalog using a search radius of $5^{\prime\prime}$, we created false lists of optical coordinates by adding $1^\circ$, $2^\circ$, and~$3^\circ$ to the right ascension coordinates in the BL1 and NL1 catalogs and~checked for FIRST detections with the same $5^{\prime\prime}$ search radius. (A similar method was employed by, e.g.,~\cite{Arsenov2025,Orosz}, when cross-matching the Quaia~\cite{quaia_cat} and the VLASS and~the FIRST and SDSS catalogs, respectively). Since we do not expect radio emission at these ``fake'' optical positions, the~found FIRST objects would indicate the rate of false detections. We found false detection rates of (0.04--0.08)\%; thus, at most, two mistakenly identified BL1 FIRST sources and one NL1 FIRST source are expected.

As a next step, we looked for counterparts of the FIRST-detected objects in the VLASS catalogs. We used a search radius of $1.5^{\prime\prime}$ around the optical positions listed in the catalogs. We used the publicly available VLASS quick-look catalogs \endnote{\url {https://cirada.ca/vlasscatalogueql0} (accessed on 2 June 2026).}. These are available for the first and second observing epochs of the VLASS. We used the third and the second pipeline versions of the epoch 1 and epoch 2 data, respectively (available at the time when the analysis was performed). Following the CIRADA VLASS Catalog User Guide\endnote{version: 26 July 2023, \url{https://ws.cadc-ccda.hia-iha.nrc-cnrc.gc.ca/files/vault/cirada/tutorials/CIRADA__VLASS_catalogue_documentation_2023_june.pdf} (accessed on 10 January 2025).}, we disregarded entries with \texttt{Duplicate\_flag}~$<$ 2, and~\texttt{Quality\_flag} not equal to $0$ or $4$, and~we only used entries with \texttt{S\_code} not equal to `E'. These settings ensured the exclusion of sidelobe structures and other false detections that may appear in the VLASS~catalogs. 

For a couple of objects with multiple FIRST counterparts within the $5^{\prime\prime}$ FIRST search radius, we checked all of those FIRST positions whether they have VLASS counterparts. If~multiple VLASS sources were found (in $8$ cases), we included their summed flux densities in the final catalog. These objects are listed in Table~\ref{tab:multi} in the Appendix. At the end, among~the BL1 and the NL1 sources detected in the FIRST, we found $1883$  and $434$ counterparts in the first-epoch VLASS catalog and~$1869$ and $445$ in the second-epoch VLASS catalog. These correspond to fractions of $\sim$73\% and $\sim$60\% for the FIRST-detected BL1 and NL1 AGN, respectively. 

Due to the non-concurrent nature of the two radio surveys, source variability can influence the analysis. With~performing the same analysis using two different epochs of the VLASS (separated by one to five years in time), we can qualitatively assess the robustness of the obtained results versus the unknown source variability.

We note that our radio-counterpart search method is not sensitive to the galaxies showing bright, extended radio lobes, with~very faint central radio emission, below~the FIRST detection limit. To~investigate the extended radio emission, lower-frequency surveys, e.g.,~those conducted by the LOFAR such as the LoTSS, Ref.~\cite{lofar_dr2}, can be more relevant, as~illustrated by, e.g., Ref. \cite{varglund2025}, who reported LOFAR detections of NLS1 galaxies.

In the following, we refer to the FIRST+VLASS1 and FIRST+VLASS2 for our samples of BL1 and NL1 AGN detected in FIRST and during the VLASS epoch 1 or VLASS epoch 2 surveys, respectively. The~data tables of the FIRST- and VLASS-detected NL1 and BL1 AGN are available at the Zenodo archive via \url{https://doi.org/10.5281/zenodo.20451035} (accessed 2 June 2026). 

\section{Results} \label{sec:results}
\unskip

\subsection{Radio~Detections}

We found that $6342$ out of the unique $51,997$ BL1, and~$2707$ out of the unique $22,590$ NL1 AGN do not fall into the FIRST sky coverage. Taking the FIRST coverage into account, the~$1.4$-GHz detection rates of the objects in the NL1 and BL1 catalogs are $\sim $3.7\% and $\sim$5.6\%, respectively, at the detection limit of the FIRST, $\sim$1\,mJy. According to the statistical Z-test~\cite{ztest}, the~probability that these proportions are not different is less than $0.1\%$. 

The distributions of Seyfert 1 galaxies and type 1 quasar among the NL1 and BL1 sources that fall within the FIRST coverage are listed in Table~\ref{tab:det_rate}. Among~the FIRST-detected ones, there are $250$ and $1046$ NL1 and BL1 galaxies, respectively, with~optical $B$-band magnitudes fainter than $-23$, thus, traditionally regarded as Seyfert galaxies. Thus, the~$1.4$-GHz radio detection rates of the NLS1 and BLS1 galaxies are $1.7\%$ and $4.3\%$, respectively. The Z-test indicates that these $1.4$-GHz radio detection rates are different at a significance level of $>$99\%. Among the FIRST-detected AGN, there are $476$ and $1515$ sources brighter than $-23$ magnitude, thus, traditionally regarded as quasars. (There were a few objects without $M_B$ values in the NL1 and BL1 catalogs). Thus, the $1.4$-GHz detection rates of the NLQ1 and BLQ1 AGN are $10.1$\,\% and $7.2$\,\%, respectively.

\begin{table}[H]
\small
\caption{Number of NL1 and BL1 sources falling within the FIRST coverage and with $1.4$-GHz radio detections in the FIRST.}
    \label{tab:det_rate}
    \begin{tabularx}{\textwidth}{CCC}
    \toprule
    \textbf{Sample} & \textbf{Within the FIRST Coverage} & \textbf{FIRST Detection}\\
    \midrule
      NL1 all & $19{,}883$ & $727$ \\
      NLQ1 & $4694$ & $476$ \\
      NLS1 & $15{,}152$ & $250$ \\
      \midrule
      BL1 all & $45{,}655$ & $2567$ \\
      BLQ1  & $21{,}131$ & $1515$ \\
      BLS1 & $24{,}424$ & $1046$\\
      \bottomrule
    \end{tabularx}
    
\end{table}

\subsection{Spectral~Index} \label{sec:alpha}

We calculated the two-point spectral index between $1.4$\,GHz and $3$\,GHz for the objects in the FIRST+VLASS1 and in the FIRST+VLASS2 samples. We flagged the objects with relative flux density errors exceeding $30\%$ in either of the two radio surveys. These were only a handful of sources, $3$ and $10$ NL1 and BL1 AGN in the FIRST$+$VLASS1 sample, and~$4$ and $5$ NL1 and BL1 AGN in the FIRST$+$VLASS2 sample, respectively. We excluded them from further analysis. A~few spectral index values were larger than the theoretically possible one for the optically thick synchrotron emission, $2.5$. We excluded those $2$ and $8$ objects from the NL1 and BL1 samples, respectively, from~the following distribution comparison. (Their details are given in Appendices~\ref{sec:outlierNLS1} and \ref{sec:outlierBLS1}). Finally, we grouped the sources according to their $M_B$ into optically bright (quasar) and fainter (Seyfert galaxy) categories. There were $1$ and $7$ ($5$ in the FIRST+VLASS2) objects in the NL1 and BL1 catalogs, respectively, lacking $M_B$ values; thus, they could not be assigned to either of the two~groups. 

\begin{figure}[H]
        \includegraphics[width=0.99\textwidth, bb=10 10 860 570, clip]{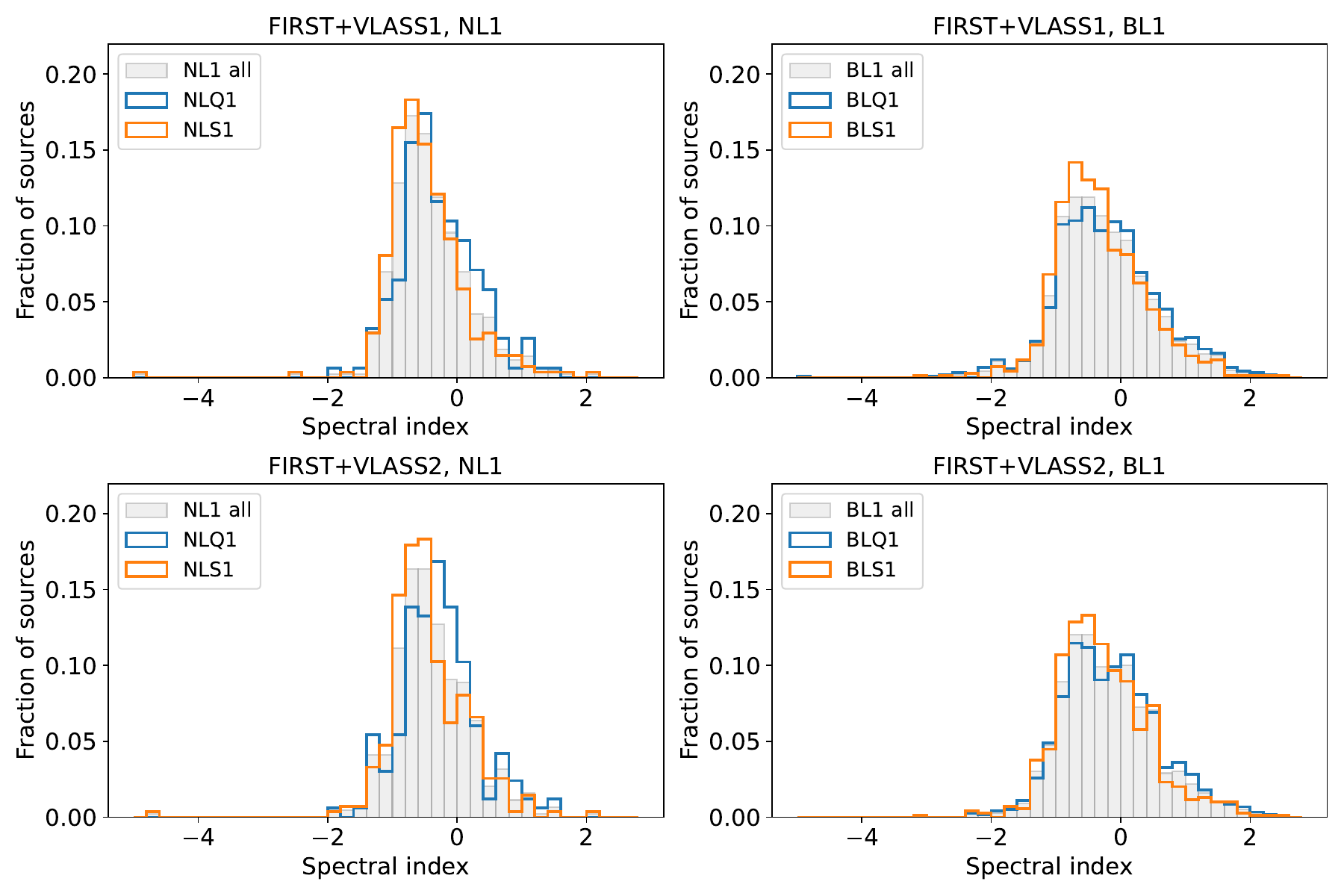}
        {\caption{Distributions of the radio spectral index values of the NL1 and BL1 samples of~\cite{paliya2024}. In~all panels, blue and yellow colors are for the respective type 1 quasar and Seyfert 1 galaxy subgroups. Grey filled histograms show the whole samples. The~top row is for the FIRST+VLASS1; the~bottom row is for the FIRST+VLASS2 samples.}
        \label{fig:alpha}}
\end{figure}

\begin{figure}[H]
    \includegraphics[width=0.99\linewidth, bb=10 10 930 390, clip]{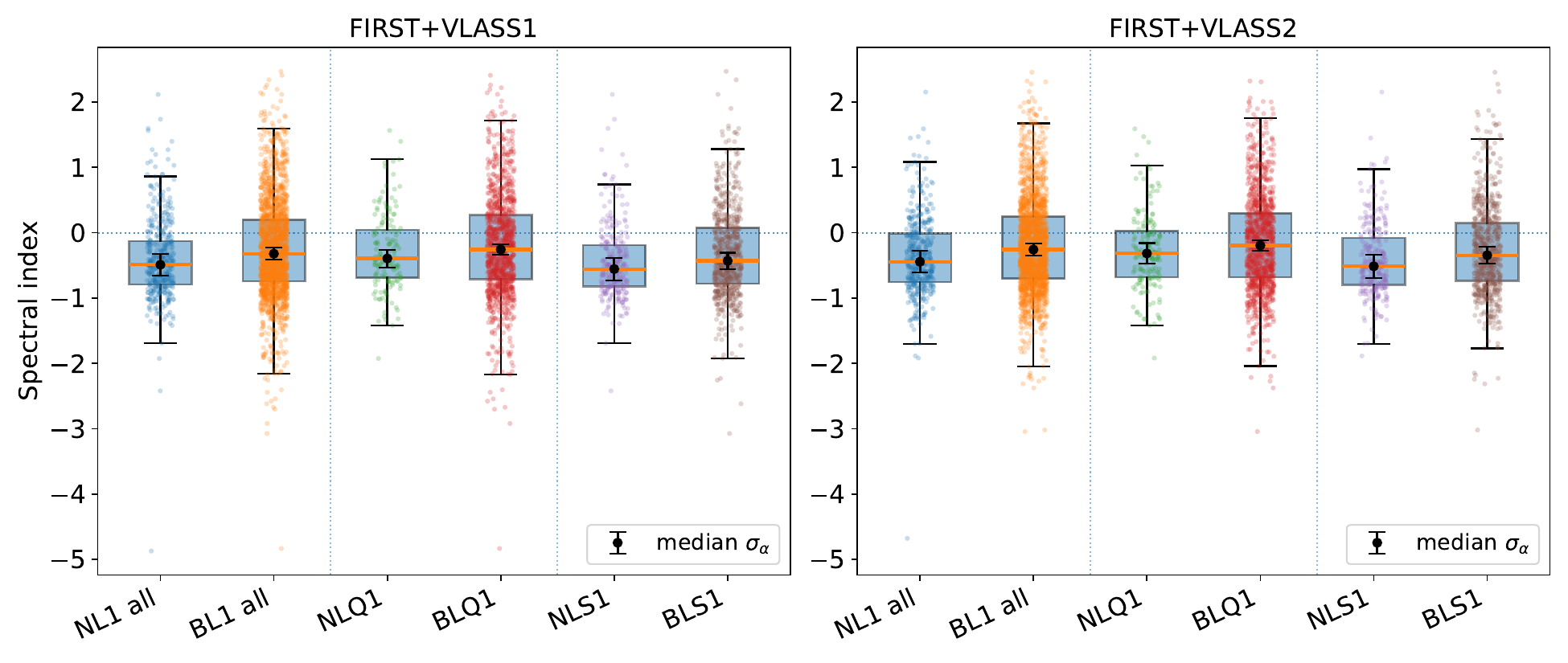}
    \caption{Boxplot representation of the spectral index distributions of the NL1 and BL1 AGN for the FIRST+VLASS1 (\textbf{left panel}) and FIRST+VLASS2 (\textbf{right panel}) samples. The~boxes extend from the first quartile, $Q_1$, to~the third quartile, $Q_3$, and~therefore, enclose the central $50$\% of the data. The~horizontal line inside each box marks the median spectral index. The~whiskers extend to the most extreme data points within $1.5$ times of the interquartile range, and~the faint points show the individual sources. The~black points and error bars indicate the median spectral index and the median of the individual spectral index uncertainties, $\sigma_\alpha$, for~the corresponding subsample. These error bars represent the typical uncertainty of individual spectral index measurements, not the uncertainty of the median. The dotted horizontal lines show the $\alpha=0$ value to guide the eye only. The vertical dotted lines separate group pairs.}
    \label{fig:alpha_boxplot}
\end{figure}

The histograms of the spectral index distributions for the FIRST+VLASS1 and FIRST+VLASS2 are shown in Figure~\ref{fig:alpha}. The~median spectral indices, the first ($Q_1$) and third quartiles ($Q_3$), and interquartile ranges ($IQR$) of the distributions are listed in Table~\ref{tab:alpha_extended}. Box diagram representations of the distributions and quartile values are shown in Figure~\ref{fig:alpha_boxplot}. The differences between the spectral indices obtained using the two epochs of the VLASS arise from source variability and possible differences between the data reduction pipelines. As illustrated by the plots and the qualitative details of the distributions, there are no major differences whether the first or the second epoch of VLASS data were used. Thus, flux density variability of individual sources, and~the different conditions of the VLASS observing epochs do not influence our results at a statistical level.

\unskip

\unskip

\begin{table}[H]
\small
\caption{Spectral index statistics of the NL1 and BL1 samples. The~table lists the number of sources, the~median spectral index, the~first and third quartiles ($Q_1$, $Q_3$), the~interquartile range ($IQR$), and~the median of the individual spectral index uncertainties ($\sigma_\alpha$) in cols. 3, 4, 5, 6, 7, and~8, respectively.}
\label{tab:alpha_extended}
\setlength{\tabcolsep}{2.0mm}{\begin{tabular}{llrccccc}
\toprule
\textbf{Data Set} & \textbf{Subsample} & \textbf{No.} & \textbf{Median \ensuremath{\boldsymbol{\alpha}}} & $\bm{Q_1~(\alpha)}$ &$\bm{Q_3~(\alpha)}$ & $\bm{IQR~(\alpha)}$ & \textbf{Median \ensuremath{\boldsymbol{\sigma_\alpha}}}\\
\midrule
\multirow[m]{6}{*}{FIRST+VLASS1} & NL1 all & $429$ & $-0.49$ & $-0.79$ & $-0.13$ & $0.67$ & $0.17$ \\
 & BL1 all & $1867$ & $-0.32$ & $-0.74$ & $0.20$ & $0.94$ & $0.09$\\
 & NLQ1 & $155$ & $-0.39$ & $-0.69$ & $0.05$ & $0.74$ & $0.14$\\
 & BLQ1 & $1169$ & $-0.26$ & $-0.71$ & $0.27$ & $0.98$ & $0.08$\\
 & NLS1 & $273$ & $-0.56$ & $-0.82$ & $-0.19$ & $0.64$ & $0.17$\\
 & BLS1 & $691$ & $-0.43$ & $-0.78$ & $0.08$ & $0.86$ & $0.12$\\
\midrule
\multirow[m]{6}{*}{FIRST+VLASS2} & NL1 all & $440$ & $-0.45$ & $-0.76$ & $-0.01$ & $0.74$ & $0.17$ \\
 & BL1 all & $1857$ & $-0.26$ & $-0.70$ & $0.25$ & $0.95$ & $0.09$ \\
 & NLQ1 & $166$ & $-0.32$ & $-0.68$ & $0.03$ & $0.71$ & $0.16$ \\
 & BLQ1 & $1160$ & $-0.20$ & $-0.68$ & $0.30$ & $0.98$ & $0.08$ \\
 & NLS1 & $273$ & $-0.51$ & $-0.80$ & $-0.08$ & $0.72$ & $0.18$ \\
 & BLS1 & $692$ & $-0.35$ & $-0.74$ & $0.15$ & $0.88$ & $0.13$ \\
\bottomrule
\end{tabular}}
\end{table}

All of the median spectral index values are $\alpha>-0.6$, formally indicating flat radio spectra. This is expected since we selected objects detected both in the FIRST and in the VLASS with mostly compact morphology; therefore, we expect a generally flat spectrum. The median value for the BLQ1 subsample corresponds to the flattest radio spectrum, and~the median value obtained for the NLS1 subsample corresponds to the steepest radio~spectrum. 

In Table~\ref{tab:alpha_extended}, we report the median of the spectral index errors of the different samples. The~median individual uncertainty is smaller than the interquartile range of the corresponding spectral index distribution. Therefore, the~observed spread of the spectral index values is not dominated by the measurement uncertainties.

\begin{figure}[H]
    \includegraphics[width=0.99\linewidth, bb= 10 10 1070 570, clip]{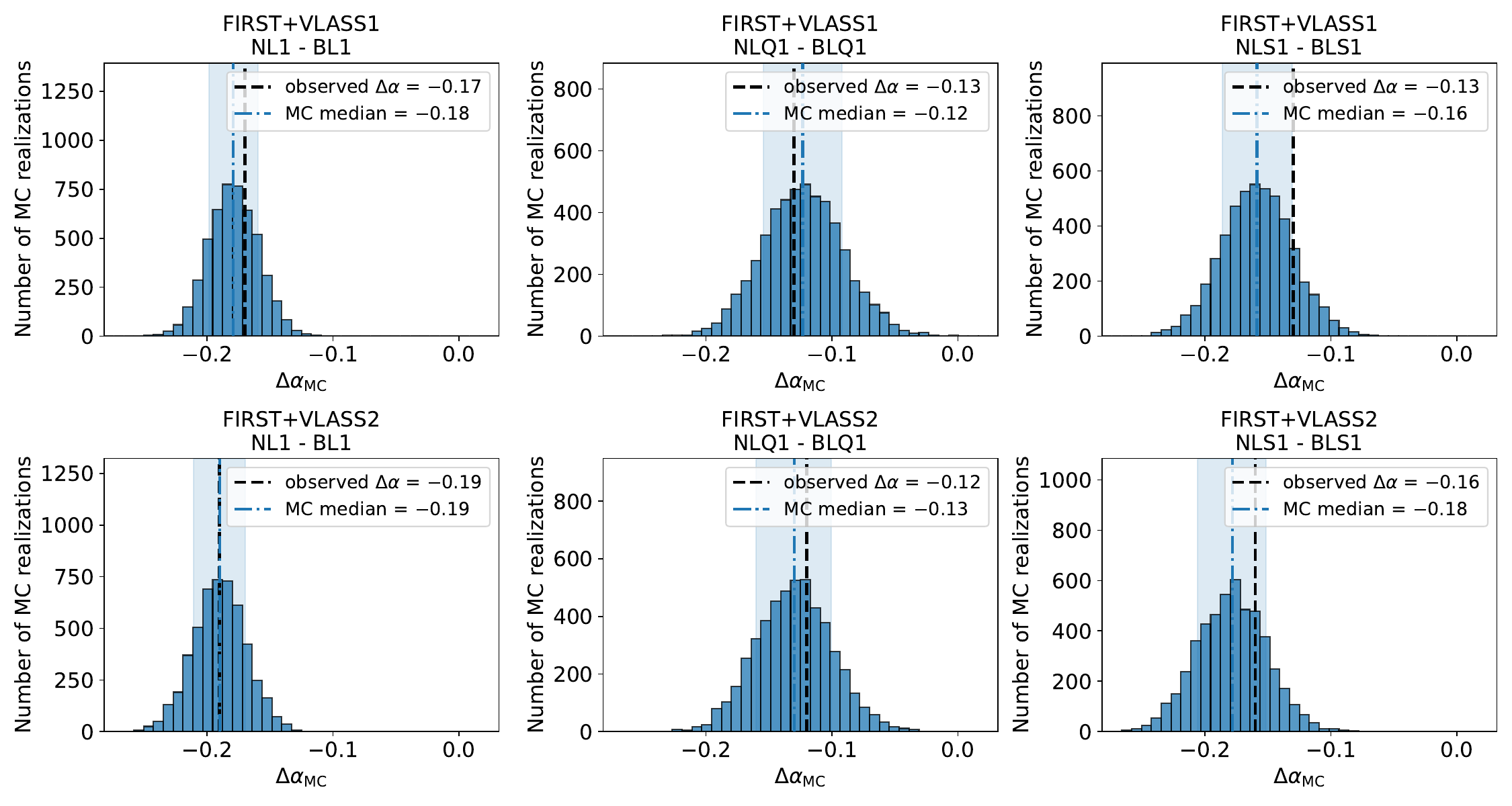}
    \caption{Monte Carlo perturbation test of the median spectral index differences between the NL1 and BL1 samples. In~each realization, every spectral index was perturbed according to its individual uncertainty, and~the median difference \(\Delta\alpha_\mathrm{MC}={\rm median}(\alpha_{\rm NL1-like})-{\rm median}(\alpha_{\rm BL1-like})\) was recomputed. The~panels show the resulting $\Delta\alpha_\mathrm{MC}$ distributions for the full samples, the~QSO subsamples, and~the Seyfert subsamples, separately for the FIRST+VLASS1 and FIRST+VLASS2 datasets. The~shaded blue regions show the 16th--84th percentile interval of the $\Delta\alpha_\mathrm{MC}$ distributions. The~black dashed line marks the observed median difference of the spectral indices, while the blue dash-dotted line marks the median of the Monte Carlo distribution. Negative values mean that the NL1 group has a steeper median spectral index than the corresponding BL1 group. The~Monte Carlo distributions remain separated from zero, indicating that the observed NL1/BL1 differences are robust against the propagated measurement uncertainties.}
    \label{fig:alpha_MC}
\end{figure}

To further test whether the observed median spectral index differences between the NL1 and BL1 samples are robust against the propagated measurement uncertainties, we performed a Monte Carlo perturbation test. In~each realization, every spectral index was perturbed according to its individual uncertainty, $\alpha_{\rm MC}\sim\mathcal{N}(\alpha,\sigma_\alpha)$, and~the difference in the median spectral indices, $\Delta\alpha_\mathrm{MC}$, was computed. We considered three comparisons: the full NL1--BL1 samples, the~type 1 quasar subsamples, NLQ1--BLQ1, and~the Seyfert 1 subsamples, NLS1--BLS1. (The tests were performed for both the FIRST+VLASS1 and FIRST+VLASS2). For all three comparisons, $\Delta\alpha_\mathrm{MC}$ was defined as the median spectral index of the NL1-like group minus that of the corresponding BL1-like group. Thus, negative values indicate that the NL1 group has a steeper median spectral index. The~Monte Carlo distributions remained centered close to the observed median spectral-index differences, $\Delta\alpha_\mathrm{obs}$, and~their central 16--84 percentile intervals do not include zero (Figure~\ref{fig:alpha_MC}). This shows that the observed NL1/BL1 spectral-index differences are robust against the propagated flux-density measurement uncertainties.

\begin{figure}[H]
        \includegraphics[width=0.99\textwidth, bb=10 10 860 570, clip]{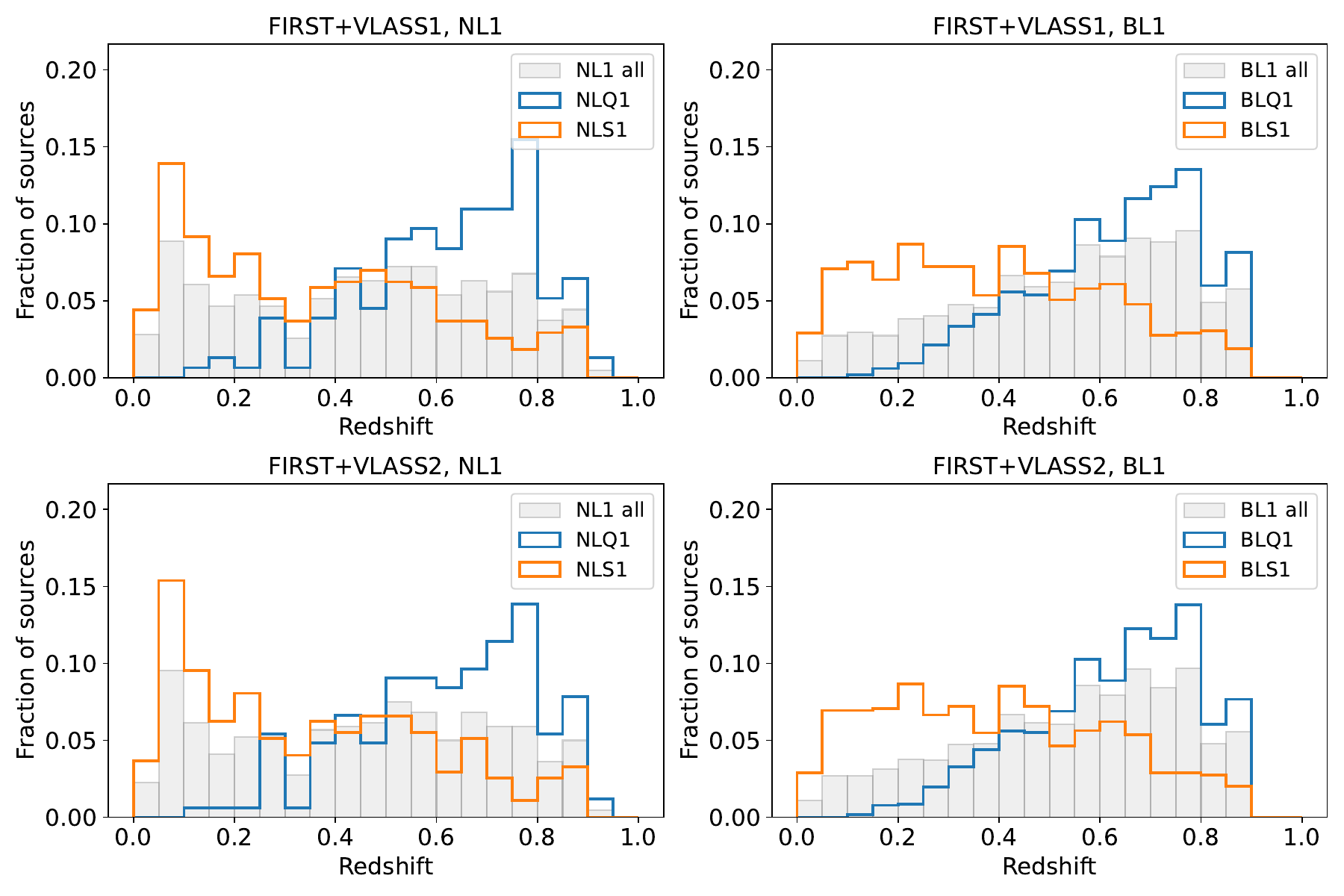}
        {\caption{Redshift distributions of the the FIRST- and VLASS-detected NL1 and BL1 samples of~\cite{paliya2024} retained for the spectral index calculation. The~details of the plots are the same as in Figure\,\ref{fig:alpha}.}
        \label{fig:radio_z}}
\end{figure}

The redshift distributions of the objects retained for the radio spectral index calculation are shown in Figure~\ref{fig:radio_z}. It is clear that the optically fainter Seyfert galaxy subgroups are not well sampled at high redshifts, where the brighter, quasar objects dominate. To test whether this can influence our results of the spectral index differences, we examined whether the redshift distributions of the NL1 and BL1 populations are statistically compatible. The~comparisons were performed separately for the full samples, the~type 1 quasar-like subsamples, and~the Seyfert 1-like subsamples, and~we used two complementary methods.

First, we compared whether using different redshift cuts on the data can provide samples with statistically compatible redshift distributions. For~each tested redshift cut, we compared the redshift distributions of the the NL1--BL1, NLQ1--BLQ1, and~NLS1--BLS1 samples (for the two VLASS epochs) via the Kolmogorov--Smirnov, e.g.,~\citep{ks} and the Anderson--Darling, AD, \citep{andersondarling} statistical tests. If~both tests failed to reject at a significance level of $5$\%, the null hypothesis that the two samples are drawn from the same redshift distribution, we regarded the redshift distributions of those two samples as compatible, not showing significant differences.
 
We found that by restricting the samples for AGN with $z>0.3$, the~redshift distributions of the pairwise compared samples no longer displayed significant differences. Then, we compared the radio spectral index distributions of the NL1-like and corresponding BL1-like groups in both the full and the $z>0.3$ samples. The~comparisons were performed separately for the NL1--BL1, NLQ1--BLQ1, and~NLS1--BLS1 sample pairs and for both VLASS epochs, using the two-sample KS and AD statistical tests. The~results are presented in detail in Appendix\,\ref{sec:stattest_z} and in Table\,\ref{tab:alpha_ksad_full_zcut}. The~full and $z>0.3$ samples give the same qualitative results.

As a second test, we applied a redshift-matched Monte Carlo resampling. Within~each redshift bin, we randomly selected the same number of AGN from each of the two compared groups and recalculated the differences of the median spectral indices in each realization. While the first method applies an empirical redshift cut, the~second one directly matches the redshift distributions, allowing us to test whether the observed NL1--BL1 spectral index differences remain after controlling for differences in the redshift distributions. Moreover, the~redshift cut removes the entire low-redshift part of the samples, and~the absence of a statistically significant difference does not imply that the remaining redshift distributions are exactly matched. The~detailed description of the Monte Carlo resampling and its results are given in Appendix \ref{sec:stattest_z}. The~differences between the subsamples remained even after the redshift resampling. 

To summarize, we found that the observed NL1/BL1 spectral index differences remain even after controlling the redshift distributions. Applying the $z>0.3$ cut does not change our main conclusions: the differences remain significant for the full NL1--BL1 samples and for the NLS1--BLS1 subsamples. In~the case of the NLQ1--BLQ1 comparisons, the~evidence for a difference is weaker for the FIRST+VLASS1 data set, when the two statistical tests gave contradictory results. However, the~spectral index difference between the subgroups NLQ1 and BLQ1 is significant when using the FIRST+VLASS2 (Table\,\ref{tab:alpha_ksad_full_zcut}).

\subsection{Radio~Loudness} \label{sec:RL}

The radio loudness parameter (RL) of an AGN is defined as the ratio of the $5$-GHz radio flux density to the $4400$\,\AA\,optical flux density~\cite{kellermann}. Paliya~et~al. (2024) \cite{paliya2024} estimated the radio loudness values from the $1.4$-GHz FIRST measurement, assuming a radio spectral index of $\alpha=-0.5$ uniformly for all sources. Since we found a broad range of actual radio spectral index values and~indications of different distributions for the different groups, we recalculated the radio loudness by using the FIRST--VLASS spectral indices, to~estimate the expected $5$-GHz flux densities of each source~individually. 

To obtain the $4400$\,\AA\,optical flux density, we calculated the $B$-band magnitudes from the SDSS-measured $g$ and $r$ magnitudes using the same formula as in~\cite{paliya2024}. We used the SDSS DR16~\cite{sdssdr16} catalog and selected only those sources where the photometry was reliable (i.e., the~`clean photometry' flag was given as $1$). 
The obtained magnitudes were corrected for galactic extinction using the extinction calculator provided by the NASA/IPAC Extragalactic Database\endnote{\url{https://ned.ipac.caltech.edu/extinction_calculator} (accessed on 2 June 2026) and the script of M. Mechtley \url{https://github.com/mmechtley/ned_extinction_calc} (accessed on 2 June 2026).}, then the magnitudes were converted to flux densities assuming a zero-point flux density of $4260$\,Jy~\cite{Bessel1979}. We assumed an optical spectral index of $-0.5$.

Similarly as was carried out for the radio spectral index investigation, we excluded those objects that have unrealistically high radio spectral indices, $\alpha>2.5$, or~have flux density measurements with relative errors $>30\%$ in either of the two radio surveys. Additionally, we wanted to minimize the effect of dust obscuration in the host galaxy, which can artificially increase the RL value. To~that end, we used the flux ratio of the H$\alpha$ to H$\beta$ emission lines. A~flux ratio value of $\gtrsim$3 implies obscuration, while values above $5$ indicate significant obscuring effect~\cite{dong-balmer-decrement-2005,dong-balmer-decrement-2008}. The~wavelength range covered by SDSS does not allow the detection of H$\alpha$ line at $z\gtrsim0.5$ and~can hinder the recognition of the line even at slightly lower redshifts as well. Therefore, we do not have information on the obscuration state of the galaxies located at $z\gtrsim0.5$. For~the RL-distribution investigation, we considered the objects that have H$\alpha$ line measurements; thus, the highest-redshift ones are not used. Among~the retained objects, we only used those that have H$\alpha$ and H$\beta$ flux density ratios not exceeding the value of $5$.

\begin{figure}[H]
    \includegraphics[width=0.99\linewidth, bb=0 10 860 580, clip]{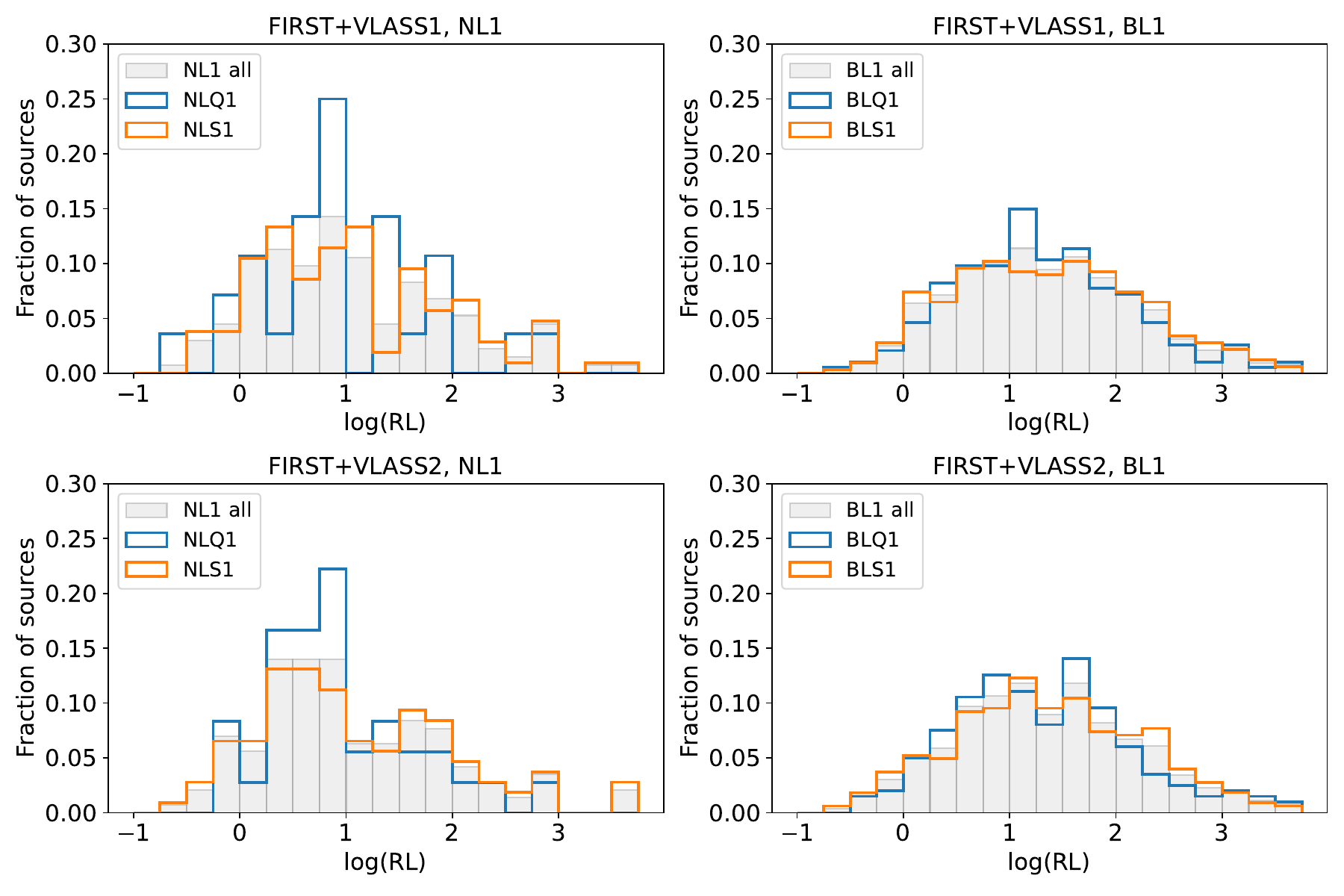}
    \caption{Distributions of the logarithm of the RL values of the NL1 and BL1 samples. The~details of the figure are the same as in Figure\,\ref{fig:alpha}.} 
    \label{fig:rl}
\end{figure}

\begin{figure}[H]
    \includegraphics[width=0.99\linewidth, bb=0 10 930 390, clip]{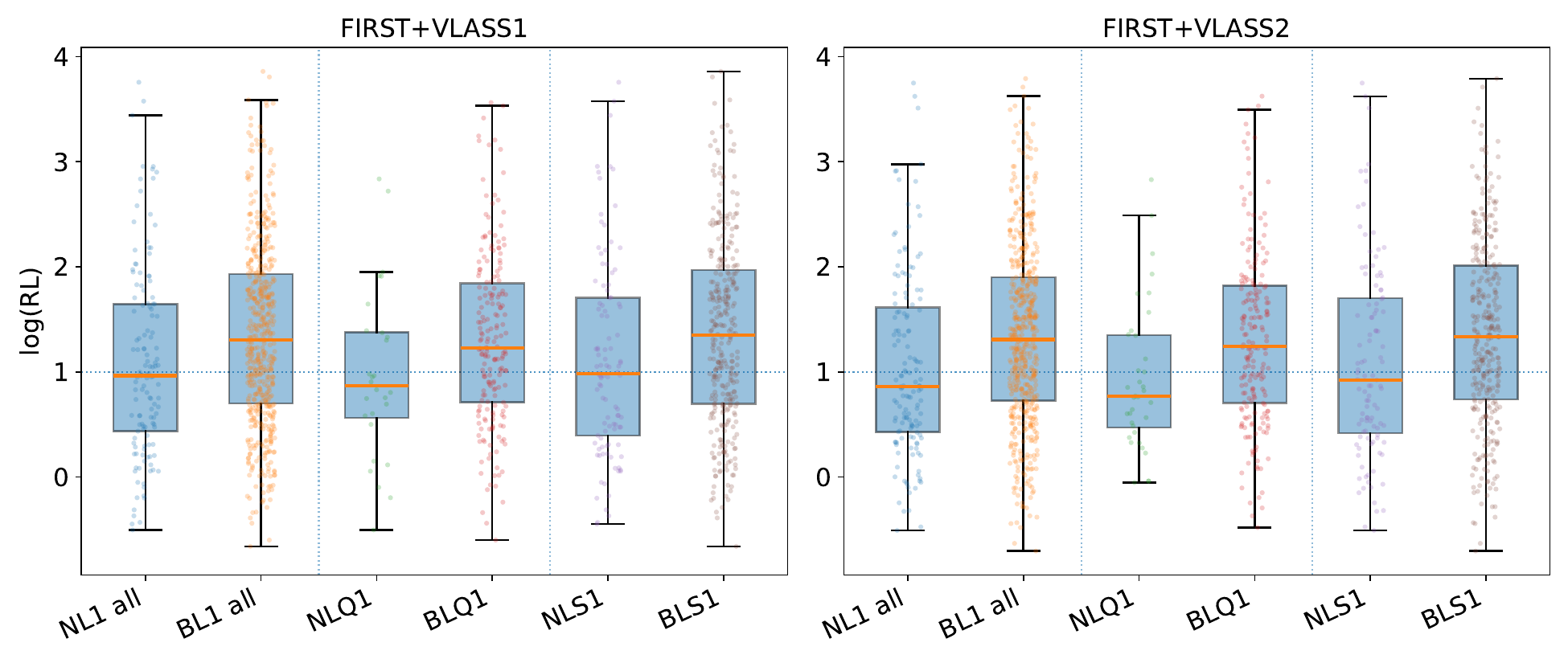}
    \caption{Boxplot representation of the radio loudness distributions of the NL1 and BL1 AGN for the FIRST+VLASS1 (\textbf{left panel}) and FIRST+VLASS2 (\textbf{right panel}) samples. The~boxes extend from the first quartile, \(Q_1\), to~the third quartile, \(Q_3\), and~therefore, enclose the central $50$\% of the data. The~horizontal line inside each box marks the median of the logarithmic radio loudness values. The~whiskers extend to the most extreme data points within $1.5$ times of the interquartile range, and~the faint points show the individual sources. The horizontal dotted line indicates the traditional division between radio-loud and radio-quiet sources~\cite{kellermann}. The vertical dotted lines separate group pairs.}
    \label{fig:rl_boxplot}
\end{figure}

The distributions of radio loudness values are shown in histogram and boxplot representations in Figures~\ref{fig:rl} and \ref{fig:rl_boxplot}, respectively. The~RL distributions of the NLQ1 are different depending on which epoch of the VLASS data is used. This can be most prominently seen in the lowest radio loudness bins in Figure~\ref{fig:rl}. It is mainly due to the small sample sizes; there are only $28$ and $36$ NLQ1 in~FIRST+VLASS1 and FIRST+VLASS2, respectively. Additionally, most of the NLQ1 that are detected only in the second VLASS epoch fall into that RL bin, with~an RL value of $\sim$3.2.

Using the KS and AD statistical tests, we compared the distributions. We obtained the same results whether we used the FIRST+VLASS1 or FIRST+VLASS2 samples. The~radio loudness of NL1 and BL1 samples cannot be described with the same distribution at a significance level of $95$\,\%. Additionally, the~radio loudness of NLS1 galaxies and the BLS1 galaxies come from different distributions, and~this is also true at the same significance levels, when comparing the NLQ1 and the BLQ1 samples. However, within~the narrow-line and broad-line samples, there are no differences in the radio loudness distributions of the optically brighter (type 1 quasar) and optically fainter (Seyfert 1 galaxy) subgroups. The~probability that BLQ1 and BLS1 come from different distributions is $\lesssim$5\%, and~similarly for the NLQ1 and NLS1 samples.

We calculated the median radio loudness values for the samples and~also the median values for the objects traditionally treated as radio-loud ones, i.e.,~$\mathrm{RL}>10$. The~resulting values are listed in Tables~\ref{tab:rl_extended} and \ref{tab:rl10_extended}. There is no substantial difference whether the RL values are calculated using the first- or the second-epoch VLASS data. The~exceptions are the median RL values of the NLQ1 subsample, showing the largest difference between the median values. This is due to the very small sample size, the~low number of NLQ1 retained for radio loudness calculations. Generally, the median RL values are larger for BL1 than NL1 AGN. But~if the radio-loud objects are considered only, the~median RL values do not differ~significantly.

Contrary to expectations that Seyfert galaxies are mostly radio-quiet (e.g., Ref.~\cite{Kharb2021} and references therein), we found that the type 1 quasar subgroups are not more radio-loud than the Seyfert 1 galaxies in our samples. We investigated whether the intermediate sources, located just at or around the traditional quasar/Seyfert galaxy division at $M_B=-23$\,mag, influence the median values. So we removed objects from the samples with $-23.5$\,mag$~< M_B<-22.5$\,mag. This, however, did not change significantly the median RL values, except~for the radio-loud NLQ1. For~them, the median RL became significantly larger, $81.2$ and $85.2$ for the FIRST+VLASS1 and FIRST+VLASS2 samples, respectively. However, such a cut further decreased the sample sizes to only $5$ sources, making the results less~constrained.

We also calculated the median radio loudness values for the datasets binned according to their absolute optical magnitudes. Since the sample sizes are small, we were able to create bins each containing $\sim$37 galaxies. This way, we had $4$ and $14$ bins for the NL1 and BL1 groups, respectively. For~the radio-loud objects, $\mathrm{RL}>10$, to~retain at least $4$ bins for the smaller NL1 sample, each bin contained $\sim$16 AGN. This way, we had $4$ and $20$ bins for the NL1 and BL1 groups, respectively. We found no obvious trends in RL with the $B$-band optical~magnitude.

\begin{table}[H]
\small
\caption{Radio loudness statistics of the NL1 and BL1 samples. The~table lists the number of sources, the~median radio loudness, the~first and third quartiles of the radio loudness distribution, and~the interquartile range calculated in logarithmic space.}
\label{tab:rl_extended}
\centering
\resizebox{\textwidth}{!}{%
\begin{tabular}{llrcccc}
\toprule
\textbf{Data Set} & \textbf{Subsample} & \textbf{No.} & \textbf{Median RL} & $\bm{Q_1}$~\textbf{(RL)} & $\bm{Q_3}$~\textbf{(RL)} & $\bm{IQR~(\log_{10}(\mathrm{RL}))}$ \\
& & & & & & \textbf{[dex]} \\
\midrule
\multirow[m]{6}{*}{FIRST+VLASS1} & NL1 all & $133$ & $9.2$ & $2.7$ & $44.2$ & $1.2$ \\
 & BL1 all & $518$ & $20.1$ & $5.0$ & $85.2$ & $1.2$ \\
 & NLQ1 & $28$ & $7.4$ & $3.6$ & $23.8$ & $0.8$ \\
 & BLQ1 & $194$ & $16.9$ & $5.1$ & $69.3$ & $1.1$ \\
 & NLS1 & $105$ & $9.7$ & $2.5$ & $50.6$ & $1.3$ \\
 & BLS1 & $324$ & $22.3$ & $5.0$ & $92.8$ & $1.3$ \\
\midrule
\multirow[m]{6}{*}{FIRST+VLASS2} & NL1 all & $143$ & $7.2$ & $2.7$ & $41.0$ & $1.2$ \\
 & BL1 all & $524$ & $20.3$ & $5.3$ & $79.3$ & $1.2$ \\
 & NLQ1 & $36$ & $5.8$ & $3.0$ & $22.3$ & $0.9$ \\
 & BLQ1 & $199$ & $17.4$ & $5.1$ & $66.2$ & $1.1$ \\
 & NLS1 & $107$ & $8.4$ & $2.6$ & $50.1$ & $1.3$ \\
 & BLS1 & $325$ & $21.7$ & $5.5$ & $102.6$ & $1.3$ \\
\bottomrule
\end{tabular}}
\end{table}
\unskip

\begin{table}[H]
\caption{Radio loudness statistics of the traditionally radio-loud (RL~$>$ 10) AGN in the NL1 and BL1 samples. The~table lists the number of sources, the~median radio loudness, the~first and third quartiles of the radio loudness distribution, and~the interquartile range calculated in logarithmic space.}
\small
\label{tab:rl10_extended}
\centering
\resizebox{\textwidth}{!}{%
\begin{tabular}{llrcccc}
\toprule
\textbf{Data Set} & \textbf{Subsample} & \textbf{No.} & \textbf{Median RL} & $\bm{Q_1}$~\textbf{(RL)} & $\bm{Q_3}$~\textbf{(RL)} & $\bm{IQR~(\log_{10}(\mathrm{RL}))}$ \\
& & & & & & \textbf{[dex]} \\
\midrule
\multirow[m]{6}{*}{FIRST+VLASS1} & NL1 all & $61$ & $51.0$ & $20.8$ & $152.1$ & $0.9$ \\
 & BL1 all & $326$ & $56.3$ & $23.7$ & $171.7$ & $0.9$ \\
 & NLQ1 & $10$ & $59.9$ & $23.8$ & $87.2$ & $0.6$ \\
 & BLQ1 & $124$ & $50.0$ & $19.7$ & $143.1$ & $0.9$ \\
 & NLS1 & $51$ & $51.0$ & $16.7$ & $162.0$ & $1.0$ \\
 & BLS1 & $202$ & $61.8$ & $27.6$ & $237.1$ & $0.9$ \\
\midrule
\multirow[m]{6}{*}{FIRST+VLASS2} & NL1 all & $61$ & $56.4$ & $24.6$ & $145.8$ & $0.8$ \\
 & BL1 all & $332$ & $53.9$ & $24.4$ & $172.6$ & $0.8$ \\
 & NLQ1 & $12$ & $45.2$ & $22.5$ & $95.2$ & $0.6$ \\
 & BLQ1 & $121$ & $50.0$ & $25.8$ & $129.0$ & $0.7$ \\
 & NLS1 & $49$ & $60.2$ & $24.6$ & $152.5$ & $0.8$ \\
 & BLS1 & $211$ & $56.6$ & $23.2$ & $209.5$ & $1.0$ \\
\bottomrule
\end{tabular}}
\end{table}

So far, we restricted the investigations to galaxies having information on their obscuration state. However, we also examined the few extremely radio-loud ($\log{(\mathrm{RL})} > 4$) NL1 and BL1 AGN, regardless of whether they have H$\alpha$ measurements available. We found that the optical spectra of three extremely radio-loud NL1 AGN are more similar to those of heavily obscured AGN. Therefore, they were removed from the NL1 sample. There are no H$\alpha$ measurements for the other two extremely radio-loud NL1 and BL1 AGN, since all of them are located at $z>0.5$. Therefore, they were not included in the RL distribution investigation. The~details of the extremely radio-loud objects are given in Appendix~\ref{sec:extremeRL}.

\subsection{Radio~Power} \label{sec:radiopower}

\begin{figure}[H]
 \includegraphics[width=0.99\linewidth, bb=10 10 860 570, clip]{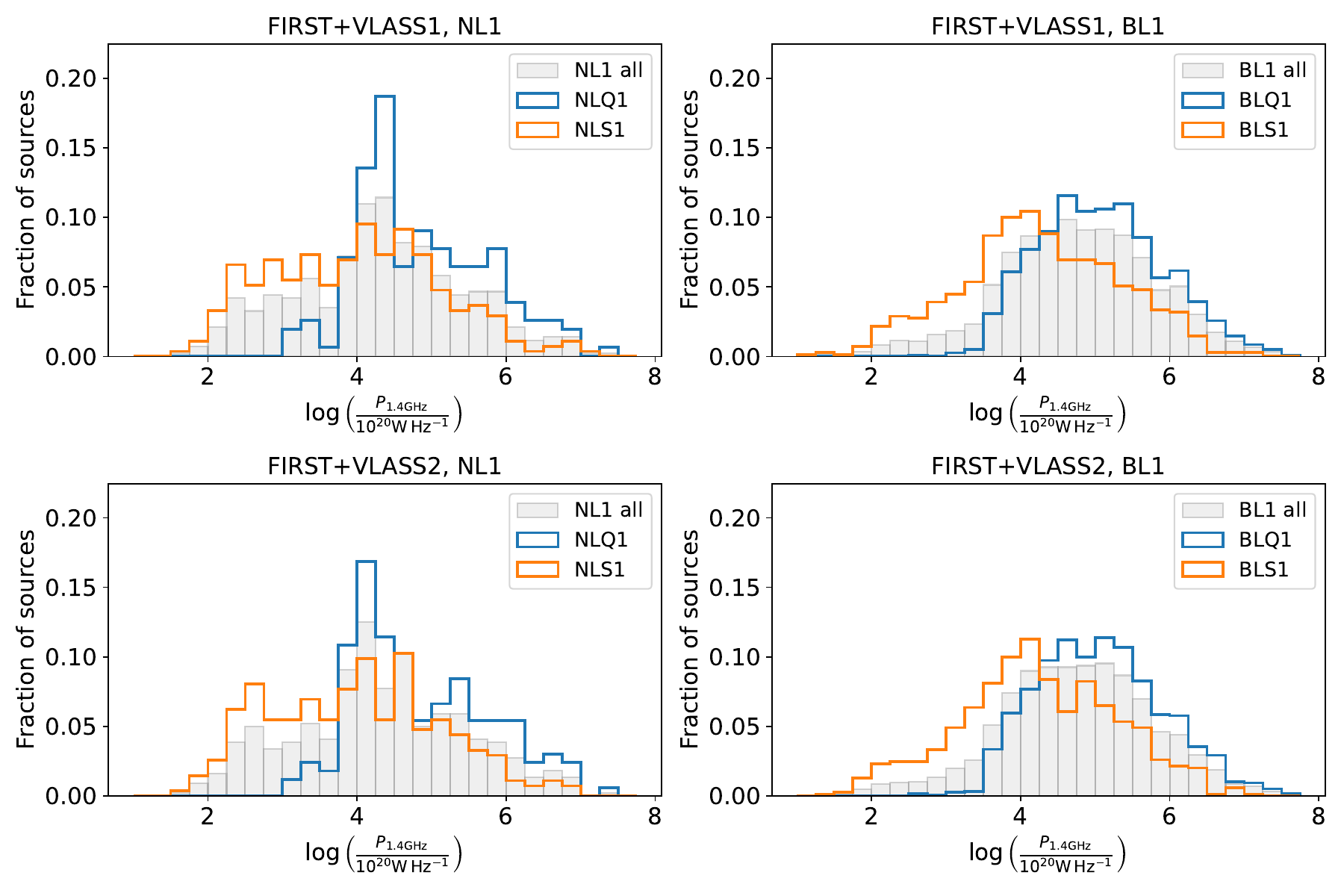}
    \caption{Distributions of the logarithm of the $1.4$-GHz radio power values in units of $10^{20}$\,W\,Hz$^{-1}$ of the NL1 and BL1 samples. The~details of the figure are the same as in Figure\,\ref{fig:alpha}.}
    \label{fig:FIRSTpower}
\end{figure}

\begin{figure}[H]
 \includegraphics[width=0.99\linewidth, bb=10 10 860 570, clip]{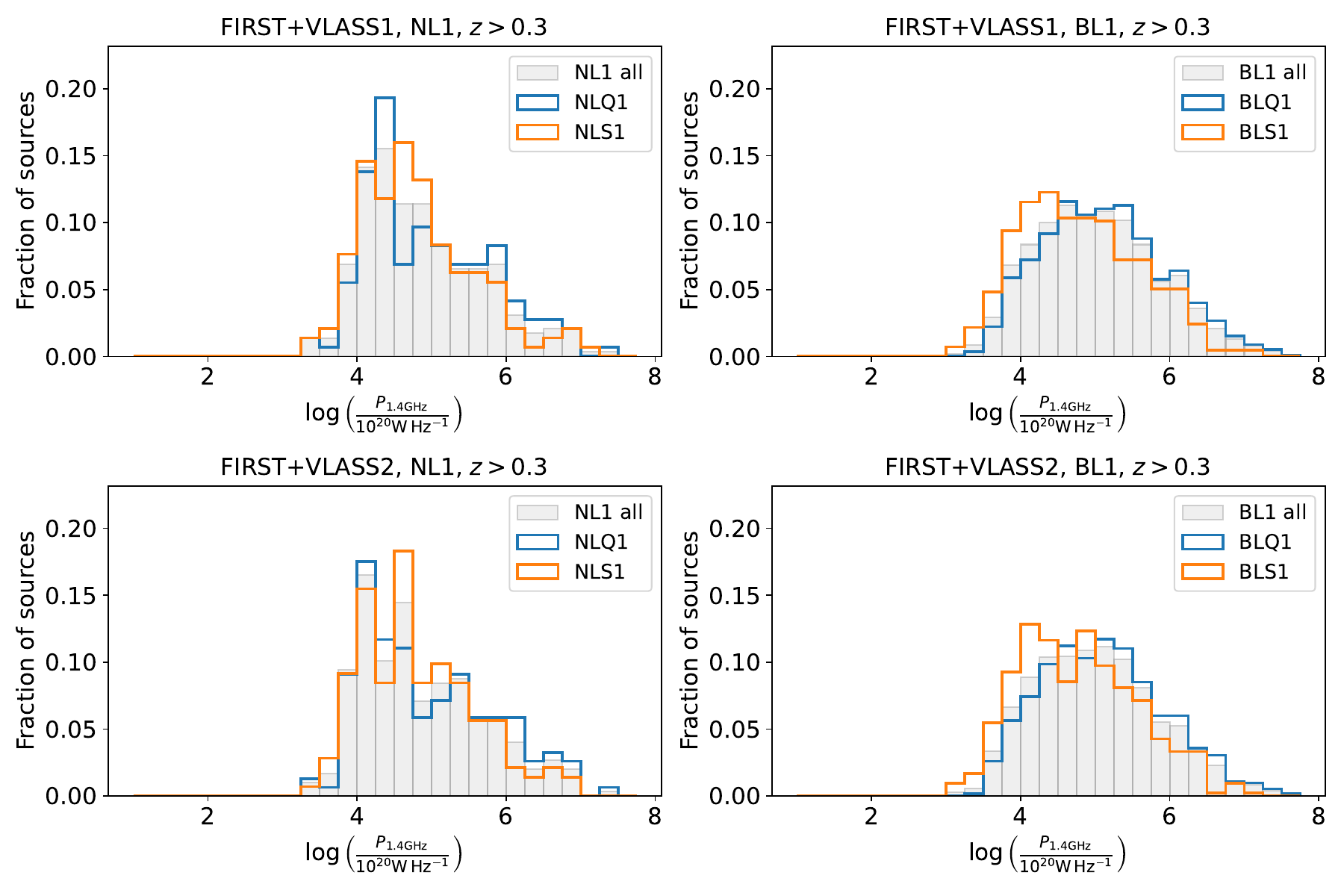}
    \caption{Distributions of the logarithm of the $1.4$-GHz radio power values in units of $10^{20}$\,W\,Hz$^{-1}$ of the NL1 and BL1 samples for sources at $z>0.3$. The~details of the figure are the same as in Figure\,\ref{fig:alpha}.}
    \label{fig:FIRSTpower_highz}
\end{figure}

We calculated the $K$-corrected $1.4$-GHz radio power, $P_{1.4\mathrm{\,GHz}}$, of~the NL1 and BL1 galaxies using their FIRST flux densities, $S_\mathrm{FIRST}$, and~the FIRST--VLASS spectral indices, $\alpha$, as 
\begin{equation}
    P_{1.4\mathrm{\,GHz}}=4\pi D_\mathrm{L}^2 S_\mathrm{FIRST} (1+z)^{-\alpha-1},
\end{equation}
where $D_\mathrm{L}$ is the luminosity distance. We disregarded the few objects with spectral indices $>2.5$ and with unreliable flux density measurements, as~before. The~distributions of $P_{1.4\mathrm{\,GHz}}$ are shown in Figures~\ref{fig:FIRSTpower} and \ref{fig:FIRSTpower_highz} for the whole samples and for the galaxies located at higher redshifts, with~$z>0.3$, respectively. This limiting redshift value was selected because the fainter Seyfert 1 galaxies are overrepresented at $z<0.3$ (see Figure~\ref{fig:radio_z}, and~Section\,\ref{sec:alpha}).

\vspace{-8pt}

\begin{figure}[H]
    \includegraphics[width=0.99\linewidth, bb=0 10 930 390, clip]{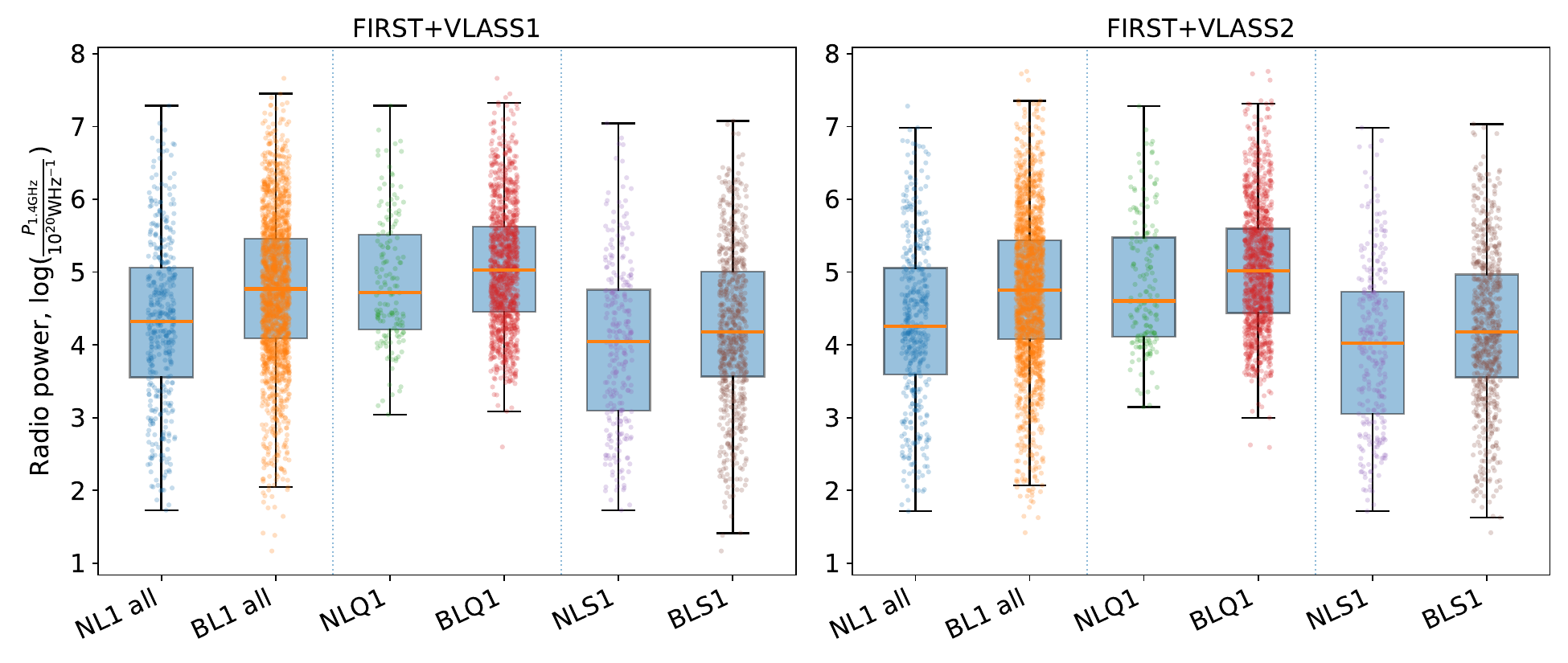}
    \caption{Boxplot representation of the FIRST radio power distributions of the NL1 and BL1 AGN. The~plot details are the same as in Figure~\ref{fig:rl_boxplot}.}
    \label{fig:rpower_box}
\end{figure}

We calculated the median radio powers quartiles and interquartile ranges for the samples, for~all objects (Table~\ref{tab:rpower_extended} and Figure~\ref{fig:rpower_box}), and for only those located at $z>0.3$ (Table~\ref{tab:rpower_highz_extended}). The~slight differences between the values obtained from the two epochs of VLASS data are most probably due to source variability. These differences 
indicate the reliability of our $P_{1.4\mathrm{\,GHz}}$ calculations. Applying the redshift cut significantly increases the median $P_{1.4\mathrm{\,GHz}}$ values for the Seyfert 1 galaxies, since this cut excludes the fainter, close-by NLS1 and BLS1 galaxies (see Figures~\ref{fig:FIRSTpower} and \ref{fig:FIRSTpower_highz}).

In all cases, type 1 quasars have higher median radio powers than Seyfert 1 galaxies. If~restricted to $z>0.3$ and using the second-epoch VLASS data, there is no difference between the median $P_{1.4\mathrm{\,GHz}}$ values of type 1 quasars and Seyfert 1 galaxies within the narrow-line sample, and~the difference is also quite small if the first-epoch VLASS data are considered. The~median $P_{1.4\mathrm{\,GHz}}$ values are similar for broad-line and narrow-line Seyfert 1 galaxies within the same redshift ranges. As~a whole, the~NL1 galaxies have lower median $P_{1.4\mathrm{\,GHz}}$ values than the BL1 galaxies.

\begin{figure}[H]
    \includegraphics[width=0.99\textwidth, bb=10 20 920 640, clip]{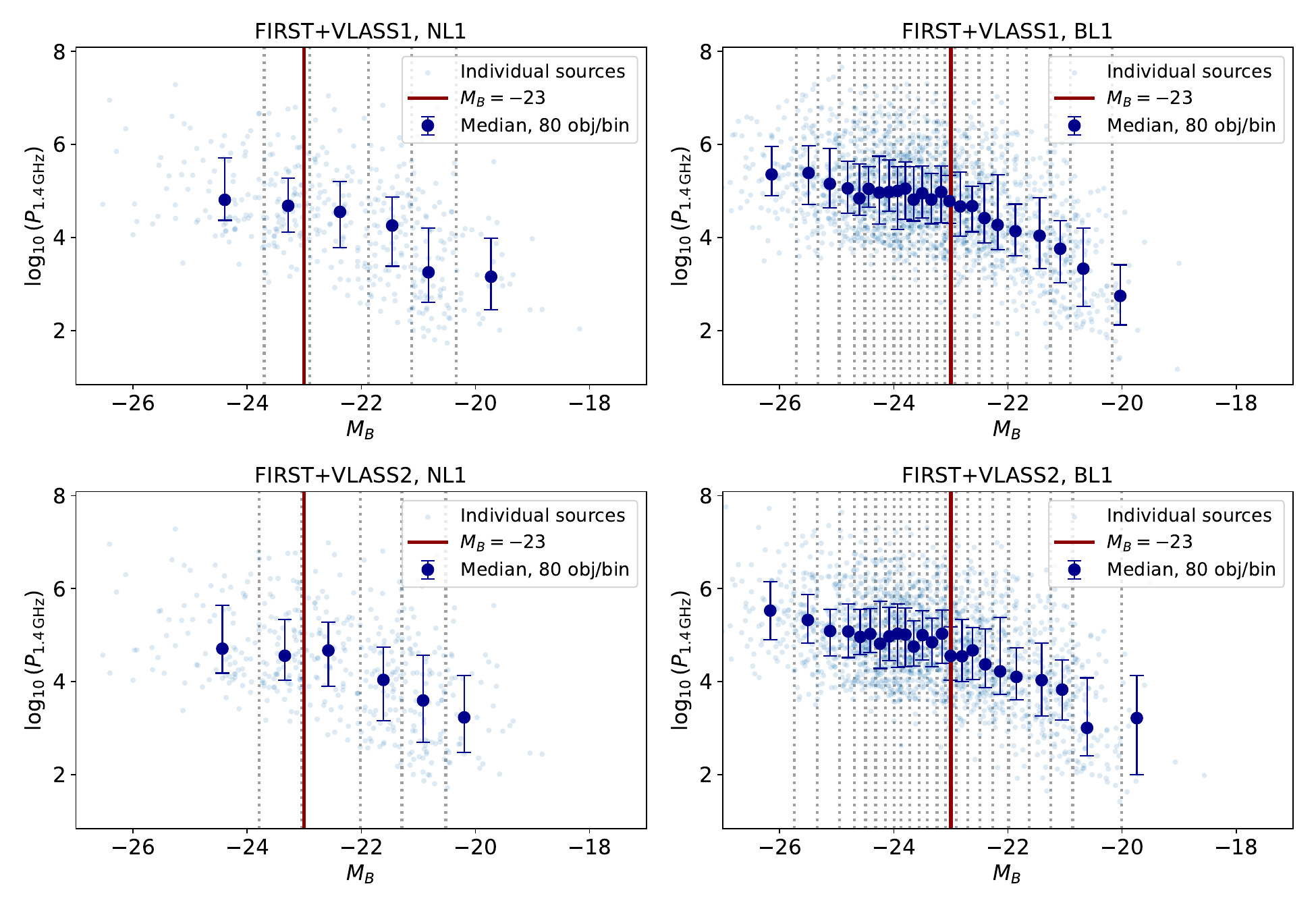}
     \caption{Optical absolute magnitude versus the logarithm of $P_{1.4\mathrm{\,GHz}}$ for the NL (\textbf{left}) and BL (\textbf{right}) samples of the FIRST+VLASS1 (\textbf{top}) and FIRST+VLASS2 (\textbf{bottom}) data. Faint points represent individual sources. Dark-blue points show the median $\log_{10}(P_{1.4\,{\rm GHz}})$ values in equal-number bins along the $M_B$ axis, with~$80$ objects per bin. The~vertical error bars show the 25--75 percentile range within each bin. Grey dotted vertical lines mark the variable-width bin boundaries, and~the red vertical line marks the QSO/Seyfert division at $M_B=-23$.}
      \label{fig:bin_FIRSTpower}
\end{figure}

To see if there is any trend in the $P_{1.4\mathrm{\,GHz}}$ values with optical absolute magnitude, we calculated the medians for different bins of $M_B$ (Figure~\ref{fig:bin_FIRSTpower}). Each bin contained $80$ AGN for the BL1 and NL1 samples. The~decreasing trends in the radio power with decreasing optical luminosity can be seen for both the broad- and narrow-line samples. (Although the sampling at fainter optical magnitudes is much sparser for the NL1 AGN).

\begin{table}[H]
\caption{Logarithmic radio-power statistics of the NL1 and BL1 samples. The~table lists the number of sources, the~median, the~first and third quartiles, and~the interquartile range for the logarithm of the normalized radio power, $P_{20} = \frac{P_{1.4\,\mathrm{GHz}}}{10^{20}\mathrm{\,W\,Hz^{-1}}}$. The~values are given for the FIRST+VLASS1 and FIRST+VLASS2 data sets above and below the horizontal line, respectively.}
\label{tab:rpower_extended}
\begin{adjustwidth}{-\extralength}{0cm}
\begin{tabularx}{\fulllength}{Lrcccc}
\toprule
  \textbf{Subsample} & \textbf{No.} & \textbf{Median $\bm{\log_{10} {\left( P_{20} \right)}}$} & $\bm{Q_1~\left[\log_{10}{\left( P_{20}\right)}\right]}$ & $\bm{Q_3~\left[\log_{10}{\left( P_{20}\right)}\right]}$ & $\bm{IQR~\left[\log_{10} {\left(P_{20}\right)} \right]}$ \textbf{[dex]}   \\
\midrule
 NL1 all & $429$ & $4.3$ & $3.55$ & $5.06$ & $1.51$ \\
  BL1 all & $1867$ & $4.8$ & $4.09$ & $5.46$ & $1.37$ \\
 NLQ1 & $155$ & $4.7$ & $4.21$ & $5.52$ & $1.30$ \\
  BLQ1 & $1169$ & $5.0$ & $4.46$ & $5.63$ & $1.17$ \\
  NLS1 & $273$ & $4.1$ & $3.10$ & $4.76$ & $1.66$ \\
  BLS1 & $691$ & $4.2$ & $3.56$ & $5.01$ & $1.44$ \\
\hline
 NL1 all & $440$ & $4.3$ & $3.59$ & $5.06$ & $1.47$ \\
  BL1 all & $1857$ & $4.8$ & $4.08$ & $5.44$ & $1.36$ \\
  NLQ1 & $166$ & $4.6$ & $4.11$ & $5.48$ & $1.36$ \\
  BLQ1 & $1160$ & $5.0$ & $4.44$ & $5.60$ & $1.16$ \\
  NLS1 & $273$ & $4.0$ & $3.05$ & $4.73$ & $1.68$ \\
  BLS1 & $692$ & $4.2$ & $3.55$ & $4.97$ & $1.41$ \\
\bottomrule
\end{tabularx}
\end{adjustwidth}
\end{table}
\unskip

\begin{table}[H]
\caption{Logarithmic radio-power statistics of the $z>0.3$ NL1 and BL1 AGN. The~details of the table are the same as in Table\,\ref{tab:rpower_extended}.}
\label{tab:rpower_highz_extended}
\begin{adjustwidth}{-\extralength}{0cm}
\begin{tabularx}{\fulllength}{Lrcccc}
\toprule
  \textbf{Subsample} & \textbf{No.} & \textbf{Median $\bm{\log_{10} {\left( P_{20} \right)}}$} & $\bm{Q_1~\left[\log_{10}{\left( P_{20}\right)}\right]}$ & $\bm{Q_3~\left[\log_{10}{\left( P_{20}\right)}\right]}$ & $\bm{IQR~\left[\log_{10} {\left(P_{20}\right)} \right]}$ \textbf{[dex]} \\
\midrule
 NL1 all & $290$ & $4.7$ & $4.28$ & $5.42$ & $1.15$ \\
  BL1 all & $1543$ & $5.0$ & $4.40$ & $5.58$ & $1.18$ \\
  NLQ1 & $145$ & $4.8$ & $4.33$ & $5.55$ & $1.22$ \\
  BLQ1 & $1124$ & $5.1$ & $4.50$ & $5.65$ & $1.15$ \\
 NLS1 & $144$ & $4.7$ & $4.20$ & $5.26$ & $1.06$ \\
 BLS1 & $416$ & $4.7$ & $4.15$ & $5.33$ & $1.18$ \\
\hline
 NL1 all & $297$ & $4.7$ & $4.18$ & $5.39$ & $1.21$ \\
  BL1 all & $1540$ & $5.0$ & $4.38$ & $5.57$ & $1.19$ \\
  NLQ1 & $154$ & $4.7$ & $4.16$ & $5.53$ & $1.36$ \\
  BLQ1 & $1116$ & $5.1$ & $4.49$ & $5.62$ & $1.13$ \\
  NLS1 & $142$ & $4.7$ & $4.21$ & $5.29$ & $1.08$ \\
  BLS1 & $421$ & $4.8$ & $4.16$ & $5.32$ & $1.17$ \\
\bottomrule
\end{tabularx}
\end{adjustwidth}
\end{table}
\unskip

Using the KS and AD statistical tests, we compared the distribution of $P_{1.4\mathrm{\,GHz}}$ values of the different subgroups. When all objects, irrespective of their redshifts, were used, the~distributions of all groups were different from each other. However, when only the $z>0.3$ objects were considered, comparing the NLS1 and BLS1 galaxies, the tests showed they came from the same distribution at a confidence level of $95$\%. As~a whole, the~$1.4$-GHz radio power distributions are different from the BL1 and NL1 samples, but~this is due to the different distributions of the optically brighter, type 1 quasar~subgroups.

We calculated the star-formation rate, SFR, from~$P_{1.4\mathrm{\,GHz}}$, assuming that all the radio flux density detected by FIRST originate from star formation, following the relation given by~\cite{Hopkins2003}
\begin{equation}
\mathrm{SFR}_\mathrm{FIRST}=\frac{P_{1.4\mathrm{\,GHz}}}{1.81 \cdot 10^{21}} \times
\begin{cases}
    1, & \text{if } P_{1.4\mathrm{\,GHz}} > 6.4 \cdot 10^{21}\\ 
    \left(0.1 + 0.9\left(\frac{P_{1.4\mathrm{\,GHz}}}{6.4 \cdot 10^{21}}\right)^{0.3} \right)^{-1}, & \text{otherwise} \\
    \end{cases}
\end{equation}
where $P_{1.4\mathrm{\,GHz}}$ is in W\,Hz$^{-1}$, and~SFR is obtained in M$_\odot$\,yr$^{-1}$. The~median SFRs for both samples are $\sim$$10^{3}\,\mathrm{M}_\odot \mathrm{\,yr}^{-1}$, irrespective of whether the data of the first or second VLASS epochs are used. Thus, to~explain the $1.4$-GHz radio power, SFR values exceeding $1000$\,$\mathrm{M}_\odot \mathrm{\,yr}^{-1}$ would be required for half of the sources in both samples. These values are unrealistically high and indicate that jetted AGN have to contribute to the radio emission in a large fraction of sources in these~samples.

\subsection{Radio~Compactness} \label{sec:comp}

To investigate the compactness of the radio emission, $C_\mathrm{FIRST}$, we calculated the peak intensity to total flux density ratios based on the FIRST data. This ratio should fall between $0$ and $1$\,beam$^{-1}$, the~latter representing point-like, unresolved emission within the given survey. However, in~reality, ratios often exceed $1$, due to various survey-related effects, e.g.,~source structure not adequately described by an elliptical Gaussian brightness distribution function~\cite{first_white}.

\begin{figure}[H]
\includegraphics[width=0.82\textwidth, bb=10 10 460 340, clip]{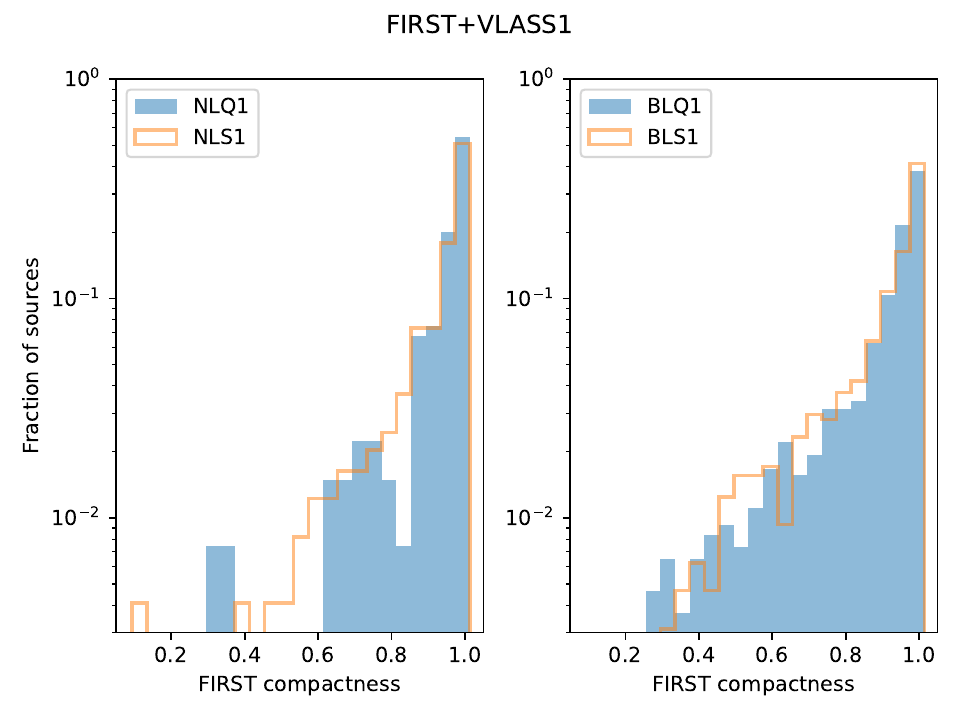}
     \caption{Distribution of FIRST compactness values for the NL1 (\textbf{left}) and BL1 (\textbf{right}) samples detected in the FIRST and in the first epoch of VLASS. Filled blue histograms represent the optically brighter type 1 quasars, orange lines the optically fainter~Seyfert 1 galaxies.}
      \label{fig:FIRSTcompact}
\end{figure}

When calculating $C_\mathrm{FIRST}$, we took into account the errors of the flux density, and we used $3$ times the rms value as an error for the peak intensity. If~$C_\mathrm{FIRST} > 1$~\,beam$^{-1}$, but~its error allowed it to be equal to $1$, we set its value to $1$. Otherwise, we did not use that object for the following compactness analysis. We also excluded the objects with multiple FIRST detections, and~similarly to before, objects with spectral index exceeding $2.5$ and~those having excessive flux density errors. The~obtained distributions are shown in Figure~\ref{fig:FIRSTcompact}. Most of the galaxies in both samples have very compact radio emission in FIRST. The~median compactness values are $>$0.95 for all samples. Only $\sim$11\% of NL1 and $\sim$20\% of BL1 galaxies have compactness values below $0.8$. This is not surprising, since the samples contain sources detected both in FIRST and in the higher-resolution VLASS survey, thus, they are expected to contain compact radio-emitting~features.

The KS and AD statistical tests showed that the compactness distributions are different from the BL1 and NL1 samples. When comparing the optically brighter quasar subgroups of the broad- and narrow-lines samples, the~tests showed different distributions, as well as when the distributions of the optically fainter NLS1 and BLS1 subgroups are~compared.

We investigated the source compactness using the second-epoch VLASS data as well, $C_\mathrm{VLASS}$. (We decided not to use the first-epoch VLASS data, since the peak intensities could be significantly underestimated at different levels at different observing times within this epoch, according to~\cite{vlass_lacy}). We followed the same approach as for the FIRST data: we disregarded sources if their compactness values, when taking into account the errors of the integrated flux density and peak intensity, were still exceeding unity, $C_\mathrm{VLASS} >1$\,beam$^{-1}$. We found that the median VLASS compactness is slightly lower $\approx0.9$ for both the BL1 and NL1 samples than that of $C_\mathrm{FIRST}$ and~slightly larger fractions, $\sim$30\% of the NL and $\sim$25\% of the BL sources have $C_\mathrm{VLASS} \lesssim 0.8$\,beam$^{-1}$. This shows that some sources point-like in FIRST have a slightly more resolved structure in the high-resolution VLASS observations; however, the majority of the objects remained unresolved.

\section{Discussion} \label{sec:discussion}
\unskip

\subsection{Origin of the Radio~Emission}

The FIRST detection rate is larger for the BL1 sources ($\sim$5.6\%) than for the NL1 sources ($\sim$3.7\%), and~a similar difference is obtained comparing the radio detection rate of BLS1 to NLS1 galaxies. Recently,~Ref.~\cite{varglund2025} investigated the radio properties of $\sim$4000 NLS1 galaxies based on the sample of~\cite{Rakshit2017}, which largely overlaps with the sample of~\cite{paliya2024}. Varglund~et~al. (2025) \cite{varglund2025} found a higher FIRST detection rate, $\sim$7.6\%, for~their NLS1 sample, using also $5^{\prime\prime}$ positional matching radius. Arsenov~et~al. (2025) \cite{Arsenov2025} investigated the radio-detection statistics of quasars by cross-matching the optical Quaia~\cite{quaia_cat} and the VLASS catalogs. They found a similar detection rate of $\sim$4\% of their~sample. 

We found that more than half of the FIRST-detected AGN were detected by VLASS ($\sim$60\% of FIRST-detected NL1 AGN and $\sim$73\% of FIRST-detected BL1 AGN). Using from these samples only those galaxies where optical spectral data do not indicate significant optical obscuration, we found that $\sim$63\% and $\sim$45\% of the radio-emitting BL1 and NL1 AGN are radio-loud ($\mathrm{RL}>10$), respectively. This is consistent with previous results of authors who reported a significant dearth of radio-loud narrow-line type 1 quasars (e.g.,~\cite{Komossa2006, Rakshit2017}) compared to broad-line type 1 quasars. The~radio-loud fractions of the BLQ1 and the optically fainter BLS1 galaxies are similarly $\sim$(60--64)\%. However, the~radio-loud fractions of NLQ1 are lower, $\sim$35\% than those of the NLS1 galaxies, $\sim$47\%.

This also means that surprisingly, we find a large fraction of the Seyfert 1 galaxy subpopulations of the BL1 and NL1 AGN  to be formally radio-loud as well, even when just focusing on systems that are known not to be very strongly reddened (and also excluding systems at high redshifts, where reddening information is lacking). This result is unexpected at first glance, given that Seyfert galaxies rarely have bright largely extended radio emission/halos, and~given that (broad-line) Seyfert galaxies are rarely of blazar type (i.e., with~relativistic jets directed at a small angle to the line of sight, thus, experiencing significant relativistic beaming); blazars are mostly found in quasars. Both bright large-scale radio emission or beaming could have contributed in enhancing their radio emission into the radio-loud regime. We now discuss other scenarios that could enhance the radio emission of Seyfert galaxies into the radio-loud~regime.

First, the~original sample of~\cite{paliya2024} is based on optical selection, including also the very faint optical populations, while the radio counterparts were selected using the FIRST with a survey threshold of $\sim$1\,mJy~\cite{Helfand_first}, and~the cross-matched sample was further constrained by selecting objects detected also in the higher-frequency VLASS observations. Thus, the~radio selection is biased towards the relatively compact and radio-bright~populations. 

Second, the~RL can be artificially enhanced by the inclusion of sources experiencing significant host extinction. We tried to mitigate this effect by including only objects with flux density ratios of their H$\alpha$ and H$\beta$ not exceeding $5$. Nevertheless, these objects can still suffer from host extinction; ideally, the limiting ratio should have been set to $3$. We also explored this possibility; however, such a cut severely reduced the sample sizes, e.g.,~only three NLQ1 sources remained, with~one of them being radio-loud. Additionally, the~results remained the same with~optically fainter objects, the~Seyfert 1 galaxy subgroups being more radio-loud in both samples.  Furthermore, in~general, type 1 AGN are not expected to be severely reddened; according to the unified model~\cite{Antonucci1993}, they are seen near~pole-on.

Third, starburst-related radio emission can contribute or even dominate in radio-loud systems. Clear indication of AGN-related radio emission can come from milliarcsecond- (mas-) scale resolution very long baseline interferometry (VLBI) observations. If~a compact radio feature with a $1.4$-GHz radio power exceeding $2 \times 10^{21}$\,W\,Hz$^{-1}$ is detected, it must be related to nuclear activity~\cite{Kewley2000}. This means that if a radio source located at $z>0.1$ is detected with VLBI, its radio emission must come from the AGN since this compact region cannot host enough star-formation-related radio-emitting sources to provide the energy output \cite{Middelberg2011}. 

Since a VLBI array, due to its narrow field of view, is not a survey instrument, we used the findings of the Deep Extragalactic VLBI--Optical Survey (DEVOS,~\cite{devos2}) to estimate the probable VLBI detection rates. According to DEVOS, $\sim$85\% of FIRST-detected unresolved sources with integrated flux densities $>20$\,mJy, that have a `star' type (i.e., unresolved) optical counterpart in SDSS can be detected by VLBI. There are $343$ BL1 in the FIRST+VLASS1, $345$ BL1 in the FIRST+VLASS2, and~$52$ NL1 AGN fulfilling these criteria with FIRST compactness values $\gtrsim 0.8$. According to the finding of DEVOS, most of these objects, $\sim$292 BL1 and $\sim$42 NL1 sources, are detectable with VLBI technique. Since all but one (a BL1 AGN) are located at redshifts exceeding $0.1$, we expect that the radio emission of these objects originate from the central AGN, according to~\cite{Middelberg2011}. Thus, as~a conservative lower limit (since VLBI non-detection does not exclude the possibility that a radio-emitting AGN can reside in the system), we can hypothesize that at least $\sim$10\% and $\sim$16\% of NL1 and BL1 AGN detected both in the FIRST and VLASS must contain radio-emitting~AGN. 

Using a different approach, one can compare the SFRs derived from the radio and infrared measurements to assess whether the detected radio emission can be explained solely by star formation or a radio-emitting AGN has to contribute. The~high median SFR$_\mathrm{FIRST}$ values for all subsamples already show that half of the objects in each subsample have SFRs exceeding $\sim$$10^3\mathrm{\,M}_\odot\mathrm{\,yr}^{-1}$ (see Section\,\ref{sec:radiopower}). Thus, it is highly probable that at least half of the sources of each subsample must contain a radio-emitting AGN to eliminate the need for unrealistically energetic star formation. Cluver~et~al. (2017) \cite{Cluver2017} provide relationships between the infrared luminosities obtained via measurements in the 12~$\upmu\mathrm{m}$ (W3) and 24~$\upmu\mathrm{m}$ (W4) bands of WISE \cite{wise}. They conclude that the tight relation between the W3 luminosity and total infrared luminosity can be used to disentangle star-formation- and AGN-related radio~emission.

\begin{figure}[H]           
\includegraphics[width=0.95\textwidth, bb=10 10 460 340, clip]{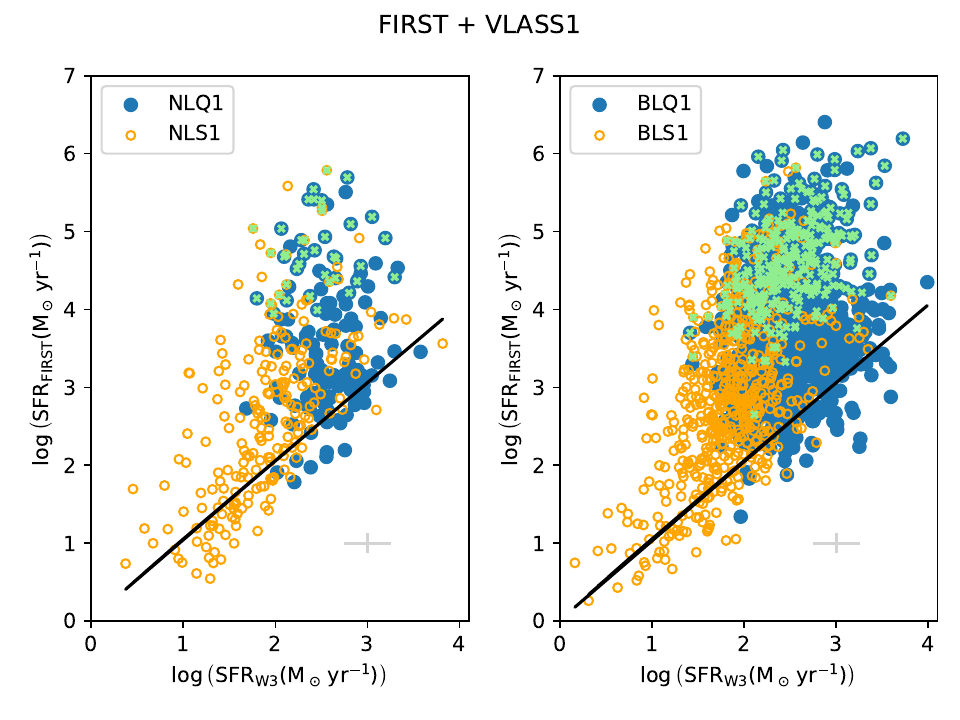}
        {\caption{\textls[13]{Star formation rates calculated from the $1.4$-GHz radio power and from the 12~$\upmu\mathrm{m}$ luminosity. Small light green symbols show the galaxies classified as `star' in SDSS, having a flux density $>20$\,mJy in FIRST and a compactness value $>0.8$ in FIRST. According to DEVOS~\cite{devos2}, these objects have a high probability to be detectable by VLBI. The black solid lines show the relationship between the SFR calculated from the $1.4$-GHz radio emission and from the 12~$\upmu\mathrm{m}$ luminosity given in~\cite{Cluver2017}. Data points are from the FIRST+VLASS1 sample. Light grey error-bars in the lower right corners illustrate the typical errors of the samples.}}
           \label{fig:sfr_compare}}
\end{figure}

We cross-matched the FIRST--VLASS samples with the AllWISE catalog~\cite{AllWISE}, and~found infrared counterparts with significant detection in the W3 band, with~SNR~$>$ 3, for~$395$ and $407$ NL1 and $1770$ and $1754$ BL1 AGN using the first- and second-epoch VLASS data, respectively. Applying Equation~(4) of~\cite{Cluver2017}, we calculated the SFR in M$_\odot$\,yr$^{-1}$ from the W3 luminosities ($\nu L_\nu$) for these galaxies as
\begin{equation}
 \log{\mathrm{SFR}_\mathrm{W3}} = 0.889 \log({\nu L_\nu})-7.76,
\end{equation}
where ($\nu L_\nu$) is in units of L$_\odot$. We used the lower three bands of WISE, 3.4~$\upmu\mathrm{m}$, 4.3~$\upmu\mathrm{m}$, and~12~$\upmu\mathrm{m}$, to~estimate the infrared spectral index and~used that for the $K$-correction in the luminosity calculation. The~obtained SFR values, SFR$_\mathrm{W3}$, are plotted against the SFR obtained from the FIRST data, SFR$_\mathrm{FIRST}$, in~Figure~\ref{fig:sfr_compare}, where we included only those objects that have radio spectral indices below $2.5$ and relative error of radio flux density measurements $<0.3$; thus, their FIRST radio power could be reliably calculated. There are $366$ and $1670$ such NL1 and BL1 AGN, respectively, in FIRST+VLASS1. Since there is no significant difference between FIRST+VLASS1 and FIRST+VLASS2 samples, we only show the results obtained with using the first-epoch VLASS~data. 

Cluver~et~al. (2017) \cite{Cluver2017} provide the relationship between the $\mathrm{SFR}_\mathrm{W3}$ and the SFR derived from $1.4$-GHz radio observations. This is shown by the black lines in Figure~\ref{fig:sfr_compare}. The~light green symbols indicate the galaxies most probably containing radio-emitting AGN according to the findings of~\cite{devos2}. All these galaxies are above the line of~\cite{Cluver2017}; thus, two independent methods show these objects most probably contain radio-emitting AGN. Concerning the BL1 and NL1 samples, most of the objects are above the line; nevertheless, the~fraction is lower for the NL1 AGN. Specifically, $79\%$ and $91\%$ of the NL1 and BL1 sources are above this relation, respectively, indicating that radio emission from the AGN in these galaxies has to contribute to the observed $1.4$-GHz radio power. The~fractions of sources where AGN contribute to the radio emission are also larger for the optically brighter quasars than for the fainter Seyfert galaxies, in~both the narrow-line and broad-line~samples.

We checked the median RL values for those galaxies that, according to the SFR comparison, most likely contain radio-emitting AGN. We found the same result as before, namely, NL1 AGN have lower median RL values ($\sim$17) than BL1 AGN ($\sim$30), and~Seyfert 1 galaxies within each sample have larger median RL than type 1 quasars. Thus, it is unlikely that star-formation-related radio emission caused the larger median RL values of Seyfert 1 galaxies.

\vspace{12pt}

\subsection{Spectral Index and Radio~Variability}

Concerning the FIRST--VLASS spectral indices, our median spectral index between $1.4$\,GHz and $3$\,GHz obtained for the NL1 sample, $-0.49$ and $-0.45$, is in good agreement with the results of~\cite{varglund2025}, $-0.61$ and $-0.47$, that were obtained with using the first- and second-epoch VLASS data, respectively. The~difference can arise from the fact that we only calculated spectral index for sources that are indicated to have the lowest probability of being observational artifacts in the VLASS catalog and~also did not include objects with large relative flux density errors. Varglund~et~al. (2025) \cite{varglund2025} also analyzed $144$-MHz LoTSS data~\cite{lofar_dr2} and were able to identify a group of NL1 AGN with peaked radio spectra. While we only calculated two-point radio spectral indices, we found that NL1 AGN have different, slightly steeper radio spectra than BL1 AGN, which may indicate difference in their radio-emitting morphology, such as containing slightly more extended~features. 

While comparing the FIRST and VLASS flux densities, we identified a couple of sources that brightened significantly between their FIRST and VLASS observations. Most of these AGN (one NL1 and four BL1) showed significant variability within $\sim$3\,yr across the three epochs of VLASS observations. However, one NL1 source, SDSS\,J110546.07$+$145202.4 (also remarked by~\cite{varglund2025} as an NLS1 galaxy with unusually large $1.4$\,GHz--$3$\,GHz radio spectral index) with the most pronounced brightening of a factor of $\sim$20, showed only hints of variability in the VLASS epochs and seems to remain in an elevated flux density state in the last $\sim$8\,yr~\cite{J1105_2025, Komossa2026}. Nyland~et~al. (2020) \cite{Nyland2020} conducted a somewhat similar study, looking for quasars undetected in FIRST but detected in VLASS with a peak intensity $\gtrsim 3$\,mJy\,beam$^{-1}$, thus ensuring a spectral index $>2.5$. They found $26$ sources transitioned from radio-quiet to radio-loud state within the last $\sim$20\,yr, but~they remained at roughly steady flux density levels during the VLASS epochs. According to their follow-up observation, some of the objects have peaked radio spectra, characteristic of young radio AGN~\cite{GPS_review}. They concluded that they most probably contain compact, young~jets. 


\section{Summary} \label{sec:summary}

We investigated the radio properties of a large sample of optically selected NL1 and BL1 active galaxies using the FIRST and VLASS radio catalogs. We separately investigated the first- and second-epoch VLASS data. Our main findings were robust irrespective of which epoch of the VLASS was~used.

\begin{itemize}
\item We found a $1.4$-GHz detection rate of a few percent for both the BL1 and NL1 samples. The~detection rate was lower for the narrow-line AGN than for the broad-line ones.
\item More than half of the $1.4$-GHz-detected NL1 and BL1 AGN were detected at $3$\,GHz in the VLASS. 
    \item Using the $1.4$-GHz FIRST and the $3$-GHz VLASS data, we found that on average, NL1 AGN have slightly steeper spectra compared to BL1 AGN, although~the median spectral index values obtained are still compatible with flat radio spectra for both samples. According to statistical tests, the~spectral index distributions of the NL1 and BL1 AGN are different. 
    \item According to the KS and AD tests, the~radio loudness of the NL1 and BL1 samples follow different distributions, with~lower median RL values for the NL1 AGN. 
    \item The optically fainter Seyfert 1 galaxies do not have lower median RL value than the optically brighter type 1 quasar samples within either the BL1 or the NL1 samples. We did not see any trend in radio loudness with $B$-band optical absolute magnitude.
    \item The $1.4$-GHz radio power of the NL1 and BL1 AGN detected at both $1.4$ and $3$-GHz show the expected trend with $B$-band optical absolute magnitude; the~radio power increases with increasing optical brightness. The~median $1.4$-GHz radio power was the largest for the optically brighter BLQ1 and the lowest for the optically fainter NLS1 sample.
    \item When comparing the SFRs estimated from the radio (FIRST) and infrared (WISE) data, we found that the radio emission of most of the investigated galaxies imply higher SFRs than the infrared emission, indicating that the AGN contribute to the radio emission significantly in those. When comparing the Seyfert 1 galaxies and the type 1 quasars clearly containing radio-emitting AGN, the~median RL value of the optically fainter Seyfert 1 galaxies is still higher than that of the type 1 quasars.
\end{itemize}

In general, NL1 AGN are fainter radio emitters, as~shown by both their $1.4$-GHz radio power and lower RL values. By~selecting AGN detected at both $1.4$\,GHz and $3$\,GHz, we mostly picked those objects whose radio emission dominantly originates from an AGN. Thus, the~differences seen in the radio emissions mainly arise from the differences in the central engines and not from the existence/lack of star formation in the host~galaxies. 

Surprisingly, the~median RL values of Seyfert 1 galaxies are higher than those of the type 1 quasars in both the narrow- and broad-line samples. However, the~$1.4$-GHz radio powers are higher for the type 1 quasars and show the expected decreasing trend with decreasing absolute optical magnitude. This apparent discrepancy may arise from the complexity when using the optical brightness derived from the observations of resolved galaxies, which can hinder the accurate estimation of radio~loudness.

Our results place the radio comparison of NL1 and BL1 AGN on a more uniform footing by exploiting the large parent catalogs of~\cite{paliya2024} together with the FIRST and the available catalogs of the first two VLASS epochs. The~persistence of the main differences across the two epoch-based analyses shows that these trends are unlikely to be driven by epoch-dependent flux-density variations, even though individual galaxies can display substantial radio variability. In~this sense, the~lower radio loudness values, lower $1.4$-GHz radio power, and~the slightly steeper radio spectra found for the NL1 sample most likely reflect genuine population-level differences in their AGN-related radio properties. A~natural next step will be to extend this work to a consistently processed multi-epoch framework using the third and fourth epochs of VLASS observations~\cite{vlass2025}, as well as the higher-quality SE data products of the VLASS. The~SE catalog is expected to include reliable spectral information within the frequency range of the survey, (2--4)\,GHz, and~would allow the determination of radio spectral index from simultaneous measurements. Thus, a~more direct test can be conducted to reveal the roles of radio variability, spectral evolution, and compact radio-source structure in shaping the observed NL1--BL1 differences.

\vspace{6pt} 

\authorcontributions{Conceptualization, K.\'E.G., S.K., and S.F.; data curation, K.\'E.G., S.K., and A.M.; methodology, K.\'E.G. and E.F.-D.; software, K.\'E.G., A.M., and E.F.-D.; formal analysis, K.\'E.G. and S.K.; visualization, K.\'E.G. and E.F.-D.; writing---original draft preparation, K.\'E.G. and S.K.; writing---review and editing, all authors. All authors have read and agreed to the published version of the manuscript.}

\funding{This research was supported by the Hungarian National Research, Development and Innovation Office via the excellence grant TKP2021-NKTA-64 and by HUN-REN Hungarian Research~Network.}



\dataavailability{The NL1 and BL1 data files used in this publication are available at the Zenodo archive at \url{https://doi.org/10.5281/zenodo.20451035} (accessed on 2 June 2026).} 

\acknowledgments{We thank the anonymous reviewers for providing useful comments that helped to improve our~paper. This research has made use of the VizieR catalog access tool, CDS, Strasbourg, France (DOI: 10.26093/cds/vizier). The~original description  of the VizieR service was published in~\cite{2000A&AS..143...23O}. 
We used in our work the Astrogeo VLBI FITS image database, DOI: 10.25966/kyy8-yp57, maintained by Leonid Petrov. 
This research has made use of the NASA/IPAC Extragalactic Database (NED), which is operated by the Jet Propulsion Laboratory, California Institute of Technology, under~contract with the National Aeronautics and Space Administration. Funding for the Sloan Digital Sky Survey IV has been provided by the Alfred P. Sloan Foundation, the~U.S. Department of Energy Office of Science, and~the Participating Institutions. SDSS-IV acknowledges support and resources from the Center for High Performance Computing  at the University of Utah. The~SDSS website is \url{www.sdss4.org}. SDSS-IV is managed by the Astrophysical Research Consortium for the Participating Institutions of the SDSS Collaboration including the Brazilian Participation Group, the~Carnegie Institution for Science, Carnegie Mellon University, Center for Astrophysics|Harvard and Smithsonian, the~Chilean Participation Group, the~French Participation Group, Instituto de Astrof\'isica de Canarias, The Johns Hopkins University, Kavli Institute for the Physics and Mathematics of the Universe (IPMU)/University of Tokyo, the~Korean Participation Group, Lawrence Berkeley National Laboratory, Leibniz Institut f\"ur Astrophysik Potsdam (AIP), Max-Planck-Institut f\"ur Astronomie (MPIA Heidelberg), Max-Planck-Institut f\"ur Astrophysik (MPA Garching), Max-Planck-Institut f\"ur Extraterrestrische Physik (MPE), National Astronomical Observatories of China, New Mexico State University, New York University, University of Notre Dame, Observat\'ario Nacional/MCTI, The Ohio State University, Pennsylvania State University, Shanghai Astronomical Observatory, United Kingdom Participation Group, Universidad Nacional Aut\'onoma de M\'exico, University of Arizona, University of Colorado Boulder, University of Oxford, University of Portsmouth, University of Utah, University of Virginia, University of Washington, University of Wisconsin, Vanderbilt University, and~Yale University.}

\conflictsofinterest{The authors declare no conflicts of~interest.} 



\abbreviations{Abbreviations}{
The following abbreviations are used in this manuscript:
\\
\vspace{-8pt}
\noindent 
\begin{longtable}[l]{@{}ll}
AD & Anderson--Darling\\
AGN & active galactic nuclei\\
AIPS & Astronomical Image Processing System\\
ASKAP & Australian Square Kilometre Array Pathfinder\\
BADASS & Bayesian AGN Decomposition Analysis for SDSS Spectra\\
BLR & Broad Line Region \\
BLS1 & Broad-line Seyfert 1\\
CIRADA & Canadian Initiative for Radio Astronomy Data Analysis\\
DEVOS & Deep Extragalactic VLBI--Optical Survey\\
DR & data release\\
EVN & European VLBI Network\\
FIRST & Faint Images of the Radio Sky at Twenty-Centimeters\\
FWHM & full-width at half maximum\\
IPAC & Infrared Processing and Analysis Center\\
KS & Kolmogorov--Smirnov\\ 
$\Lambda$CDM & Lambda Cold Dark Matter\\
LINER &  low-ionization nuclear emission-line region\\
LOFAR & Low-Frequency Array\\
LoTSS & Low-Frequency Array Two-metre Sky Survey\\
\multirow{2}{*}{MOJAVE} & Monitoring of Jets of Active Galactic Nuclei with Very Long Baseline \\
 &  Array Experiments\\
NASA & National Aeronautics and Space Administration\\
NLR & Narrow Line Region \\
NLS1 & Narrow-line Seyfert 1\\
NRAO & National Radio Astronomy Observatory\\
NVSS & NRAO VLA Sky Survey\\
RACS & Rapid ASKAP Continuum Survey\\
RL & radio loudness\\
SDSS & Sloan Digital Sky Survey\\
SE & single epoch\\
SF & star formation\\
SFR & star formation rate\\
SNR & signal-to-noise ratio\\
VLA & Karl G. Jansky Very Large Array\\
VLASS & VLA Sky Survey\\
VLBA & Very Long Baseline Array\\
VLBI & very long baseline interferometry\\
VSOP & VLBI Space Observatory Program\\
WISE & Wide-field Infrared Survey Explorer\\
\end{longtable}
}

\appendixtitles{yes} 
\appendixstart
\appendix

\section[\appendixname~\thesection]{Optical Galaxies with Multiple Radio Counterparts}
\label{sec:app_multi}

We give here the list of NL1 and BL1 AGN with multiple FIRST radio detections (Table~\ref{tab:multi}). In~these objects, the~FIRST flux densities given in our catalog are the summed flux densities of the FIRST source components. We also indicate whether multiple counterparts are found in the VLASS catalogs. In~those cases, the~VLASS flux densities in the catalog are the summed flux densities of the~counterparts.

\begin{table}[H]
\caption{Objects with multiple FIRST counterparts. The~first one is an NL1, the~others are all BL1 AGN.}
 \label{tab:multi}
    \setlength{\tabcolsep}{15.9mm}{\begin{tabular}{cc}
    \toprule
    \textbf{Name} & \textbf{VLASS Counterparts}\\
    \midrule
       SDSS\,J030313.02$-$001457.5 & 1 \\
SDSS\,J003626.98$+$001303.7 & 2  \\
SDSS\,J023814.10$-$062852.0 & 2 \\
SDSS\,J024500.70$-$074736.4 & 2 \\
SDSS\,J083337.36$+$221244.3 & 2 \\
SDSS\,J092307.39$+$305926.4 & 2 \\
SDSS\,J093000.17$+$250005.7 & 1 \\
SDSS\,J105026.03$+$015138.3 & 1 \\
SDSS\,J114707.57$+$503053.4 & 2 \\
SDSS\,J132103.41$+$123748.2 & 1 \\
SDSS\,J134303.59$+$521626.7 & 2 \\
SDSS\,J230545.66$-$003608.6 & 1 \\
\bottomrule
    \end{tabular}}
   \end{table}

\section{Statistical Investigations of the Samples' Redshift~Distributions}
\label{sec:stattest_z}
\vspace{-8pt}
\begin{table}[H]
\centering
\caption{Exploratory scan of the redshift-cut used to compare the redshift distributions of the NL1-like and corresponding BL1-like samples. $N_{\min}$ is the smallest group size among the six comparisons, comprising the NL1--BL1, NLQ1--BLQ1, and~NLS1--BLS1 pairs for FIRST+VLASS1 and FIRST+VLASS2. $\min(p_{\rm KS})$ and $\min(p_{\rm AD})$ are the smallest two-sample KS and AD $p$-values, respectively, among~these six comparisons. $N_{\rm KS}/6$ and $N_{\rm AD}/6$ give the numbers of comparisons for which the corresponding test does not reject the null hypothesis of identical redshift distributions at the adopted significance level of $p=0.05$. $N_{\rm both}/6$ is the number of comparisons for which neither test rejects the null hypothesis. The~first row represents the effectively uncut sample. The~bold row at $z_{\min}=0.30$ marks the lowest tested threshold for which neither test rejects the null hypothesis in any of the six comparisons.}
\label{tab:redshift_threshold_scan}
\setlength{\tabcolsep}{4.1mm}{\begin{tabular}{ccccccc}
\toprule
$\boldsymbol{z_\mathrm{cut}}$ & $\boldsymbol{N_{\min}}$ & $\boldsymbol{\min~(p_{\rm KS})}$ & $\boldsymbol{\min~(p_{\rm AD})}$ & $\boldsymbol{N_{\rm KS}/6}$ & $\boldsymbol{N_{\rm AD}/6}$ & $\boldsymbol{N_{\rm both}/6}$ \\
\midrule
$0.00$ & $155$ & $<$$10^{-4}$ & $\leq$0.001 & $2/6$ & $2/6$ & $2/6$ \\
$0.05$ & $155$ & $<$$10^{-4}$ & $\leq$0.001 & $2/6$ & $2/6$ & $2/6$ \\
$0.10$ & $155$ & $0.0001$ & $\leq$0.001 & $4/6$ & $4/6$ & $4/6$ \\
$0.15$ & $154$ & $0.0023$ & $\leq$0.001 & $4/6$ & $4/6$ & $4/6$ \\
$0.20$ & $152$ & $0.0059$ & $0.0046$ & $4/6$ & $4/6$ & $4/6$ \\
$0.25$ & $151$ & $0.0244$ & $0.0246$ & $4/6$ & $4/6$ & $4/6$ \\
$\boldsymbol{0.30}$ & $\boldsymbol{142}$ & $\boldsymbol{0.0643}$ & $\boldsymbol{0.0882}$ & $\boldsymbol{6/6}$ & $\boldsymbol{6/6}$ & $\boldsymbol{6/6}$ \\
$0.35$ & $131$ & $0.0216$ & $0.0157$ & $4/6$ & $4/6$ & $4/6$ \\
$0.40$ & $114$ & $0.0408$ & $0.0407$ & $5/6$ & $5/6$ & $5/6$ \\
$0.45$ & $99$ & $0.0733$ & $0.0544$ & $6/6$ & $6/6$ & $6/6$ \\
$0.50$ & $81$ & $0.1891$ & $0.0673$ & $6/6$ & $6/6$ & $6/6$ \\
\bottomrule
\end{tabular}}
\end{table}

We employed two methods to investigate whether our findings of the spectral index distributions of the different NL1 and BL1 samples (in Section\,\ref{sec:alpha}) are influenced by the different redshift distributions of the samples. In~the first approach, we compared the redshift distributions of the samples via the KS and AD statistical tests after applying different redshift cuts, $z_{\rm cut}$. The~cut values were chosen between $0.05$ and $0.50$, in~steps of $0.05$. For~each threshold, we compared the redshift distributions of the NL1--BL1, NLQ1--BLQ1, and~NLS1--BLS1 samples separately for the two VLASS epochs. We required both tests to give $p\geq0.05$ in all six epoch--comparison combinations. 

In Table~\ref{tab:redshift_threshold_scan}, we list the results of these comparisons. For~each $z_\mathrm{cut}$ value, we give the smallest group size, $N_\mathrm{min}$, the~lowest $p$-values of the KS and AD tests, $p_\mathrm{KS}$, $p_\mathrm{AD}$, respectively, and~the number of comparisons, $N_\mathrm{KS}$, $N_\mathrm{AD}$ where the KS/AD test failed to reject the null hypothesis (at a significance level of $p=0.05$) that the redshift of the pairs come from the same redshift distribution. In~the last column, we list the number of comparisons for which neither test rejects the null hypothesis. It can be seen that $z_\mathrm{cut}=0.3$ was the lowest tested threshold for which neither test rejects the null hypothesis in any of the six comparisons. At~this threshold, the~minimum KS and AD $p$-values were $0.0643$ and $0.0882$, respectively, and~the smallest retained group still contained $142$ objects. Although~the criterion was also satisfied at $z_{\rm cut}=0.45$ and $0.50$, these cuts retained substantially fewer objects, reducing the statistical power of the tests. Therefore, we adopted $z>0.3$ as the lowest threshold that removed the statistically detectable differences between the redshift distributions without unnecessarily reducing the sample size. The~$p$-values do not necessarily vary monotonically with $z_{\rm cut}$, because~each threshold changes both the composition and the size of the retained samples. 

\begin{table}[H]
 \caption{Comparison of the median spectral indices and the results of pairwise comparison of the spectral index distributions with the KS and AD tests for the whole and the $z>0.3$ samples. The~number of sources is given in the third column. The~median spectral indices of the two samples are listed in cols. 4 and 5. Here, index $1$ and $2$ refer to the first and second samples given in column 1. In~cols. 6 and 7, the $p$ values of the respective statistical tests are listed. In~the last two columns, we report whether the given test rejected the null hypothesis. Results are given for the FIRST+VLASS1 and FIRST+VLASS2 data sets above and below the horizontal line, respectively.}
    \label{tab:alpha_ksad_full_zcut}
    \begin{adjustwidth}{-\extralength}{0cm}
    \begin{tabularx}{\fulllength}{CLcccccll}
    \toprule
     \textbf{Comparison} & \textbf{Selection} & \ensuremath{\boldsymbol{N_1/N_2}} & \textbf{Median $\bm{\alpha_1}$} & \textbf{Median $\bm{\alpha_2}$} & $\bm{p_\mathrm{KS}}$ & $\bm{p_\mathrm{AD}}$ & \textbf{KS Reject} & \textbf{AD Reject} \\
\midrule
        \multirow[m]{2}{*}{NL1--BL1} & Full & $429/1867$ & $-0.49$ & $-0.32$ & $<$$10^{-4}$ & $\lesssim$0.001 & True & True \\
          & $z>0.3$ &  $290/1543$ & $-0.49$ & $-0.26$ & $<$$10^{-4}$ & $\lesssim$0.001 & True & True \\
           \multirow[m]{2}{*}{NLQ1--BLQ1} & Full & $155/1169$ & $-0.39$ & $-0.26$ & $0.0546$ & $0.0214$ & False & True \\
        & $z>0.3$ & $145/1124$ & $-0.39$ & $-0.23$ & $0.0733$ & $0.0284$ & False & True\\
        \multirow[m]{2}{*}{NLS1--BLS1} & Full & $273/691$ & $-0.56$ & $-0.43$ & $0.0008$ & $\lesssim$0.001 & True & True \\
         & $z<0.3$ & $144/416$ & $-0.61$ & $-0.32$ & $0.0001$ & $\lesssim$0.001 & True & True \\
        \midrule
           \multirow[m]{2}{*}{NL1--BL1} & Full & $440/1857$ & $-0.45$ & $-0.26$ & $<$$10^{-4}$ & $\lesssim$0.001 & True & True \\
          & $z>0.3$ &  $297/1540$ & $-0.41$ & $-0.22$ & $<$$10^{-4}$ & $\lesssim$0.001 & True & True \\
           \multirow[m]{2}{*}{NLQ1--BLQ1} & Full & $166/1160$ & $-0.32$ & $-0.20$ & $0.0560$ & $0.0081$ & True & True \\
        & $z>0.3$ & $154/1116$ & $-0.31$ & $-0.18$ & $0.0046$ & $0.0105$ & True & True\\
        \multirow[m]{2}{*}{NLS1--BLS1} & Full & $273/692$ & $-0.51$ & $-0.35$ & $0.0003$ & $\lesssim$0.001 & True & True \\
         & $z<0.3$ & $142/421$ & $-0.56$ & $-0.29$ & $0.0001$ & $\lesssim$0.001 & True & True \\
         \bottomrule
    \end{tabularx}
    \end{adjustwidth}
\end{table}

Then, we compared the radio spectral index distributions of the whole and the $z>0.3$ samples. We calculated the median spectral indices and~compared the spectral index distributions using the KS and AD tests for the NL1--BL1, NLQ1--BLQ1, and~NLS1--BLS1 sample pairs for both VLASS epochs. The~null hypothesis was that the two samples are drawn from the same underlying spectral index distributions. The~results are given in Table~\ref{tab:alpha_ksad_full_zcut}.

\begin{figure}[H]
\includegraphics[width=0.99\linewidth, bb= 10 10 1070 570, clip]{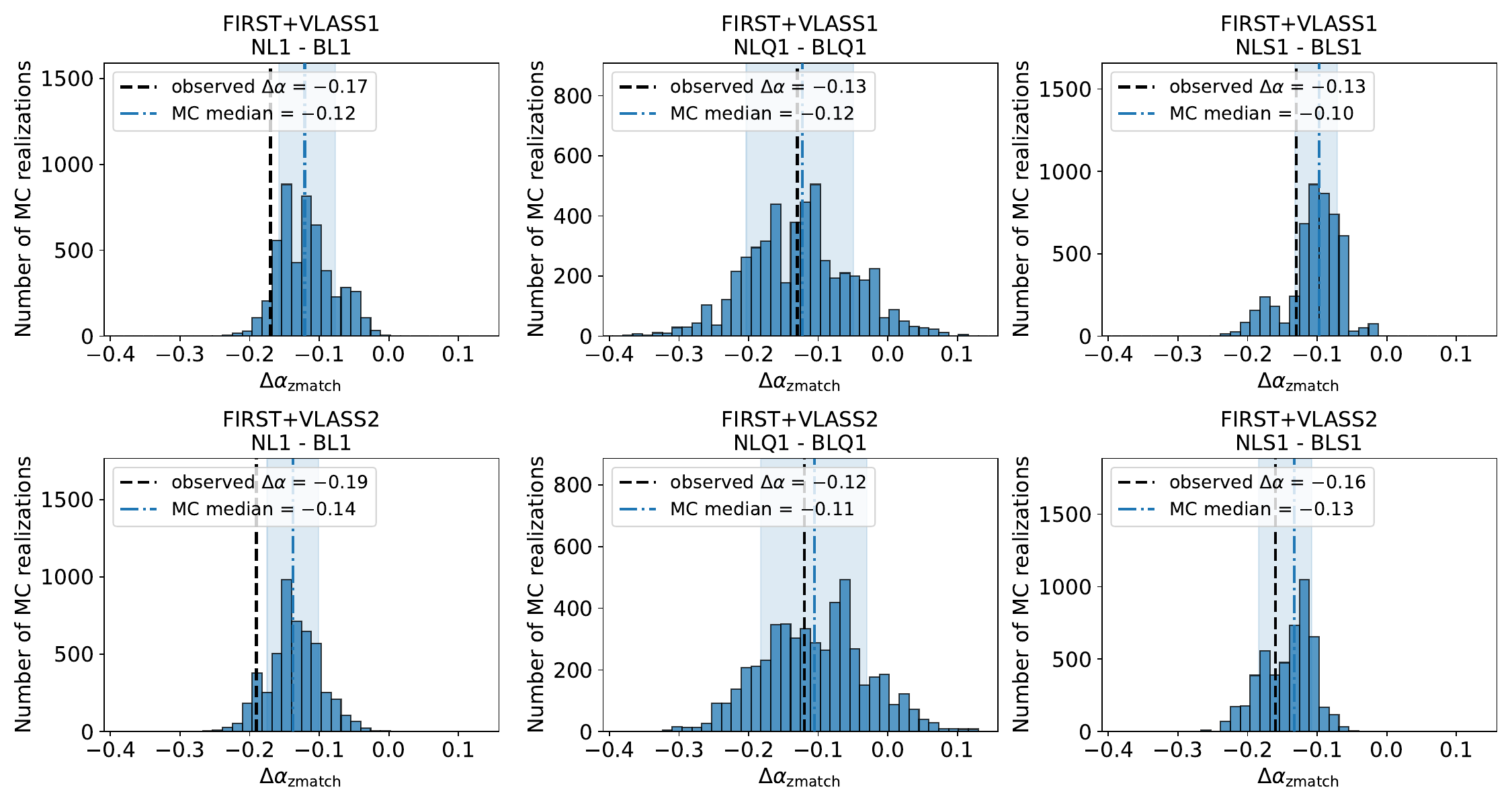} 
\caption{The redshift-matched resampling distributions of the median spectral index differences between the NL1-like and corresponding BL1-like samples. Negative values indicate that the NL1-like group has a steeper median radio spectrum than the corresponding BL1-like group. The~upper and lower rows show the FIRST+VLASS1 and FIRST+VLASS2 results, respectively, and~from left to right for the full samples, the~type 1 quasar-like subsamples, and~the Seyfert 1 like subsamples. The~black dashed lines mark the observed median differences, while the blue dash-dotted lines show the medians of the redshift-matched distributions. The~shaded regions indicate the central $16$--$84$ percentile intervals.}
\label{fig:alpha_zmatched}
\end{figure}
\unskip

\begin{table}[H]
\centering
\caption{The redshift-matched resampling test of the median spectral index differences between the NL1-like and corresponding BL1-like samples. The~tested quantity is defined as $\Delta\alpha_\mathrm{zmatch}={\rm median}(\alpha_{\rm NL1-like})-{\rm median}(\alpha_{\rm BL1-like})$, where $N_1$ and $N_2$ are the original sample sizes, and $N_{\rm match}$ is the number of matched objects in each group. The~quoted intervals correspond to the central $16\text{--}84$ percentiles of the redshift-matched distributions. Results are given for the FIRST+VLASS1 and FIRST+VLASS2 data sets above and below the horizontal line, respectively.}
\label{tab:zmatched_nl_bl_delta}
\setlength{\tabcolsep}{3mm}{\begin{tabular}{crrrccc}
\toprule
 \textbf{Comparison} & $\bm{N_1}$ & $\bm{N_2}$ & $\bm{N_{\rm match}}$ & $\bm{\Delta\alpha_{\rm obs}}$  & $\bm{\Delta\alpha_{\rm zmatch,med}}$ & \textbf{[16--84\%]}\\
\midrule
 NL1--BL1 & 429 & 1867 & 427 & $-0.17$ & $-0.12$ & $\left[-0.158,-0.077\right]$ \\
  NLQ1--BLQ1 & 155 & 1169 & 153 & $-0.13$ & $-0.12$ & $\left[-0.203,-0.049\right]$ \\
  NLS1--BLS1 & 273 & 691 & 273 & $-0.13$ & $-0.10$ & $\left[-0.132,-0.071\right]$ \\
\midrule
 NL1--BL1 & 440 & 1857 & 438 & $-0.19$ & $-0.14$ & $\left[-0.175,-0.101\right]$ \\
  NLQ1--BLQ1 & 166 & 1160 & 164 & $-0.12$ & $-0.11$ & $\left[-0.183,-0.030\right]$ \\
 NLS1--BLS1 & 273 & 692 & 273 & $-0.16$ & $-0.13$ & $\left[-0.184,-0.107\right]$ \\
\bottomrule
\end{tabular}}
\end{table}

In the second approach, to~control more directly the differences in the redshift distributions, we performed a redshift-matched Monte Carlo resampling analysis for the NL1--BL1, NLQ1--BLQ1, and~NLS1--BLS1 comparisons in both VLASS epochs. We divided the samples into redshift bins with a width of $\Delta z=0.10$. In~each redshift bin containing objects from both groups, we randomly selected the same number of sources from the two groups,
\begin{equation}  
N_{\rm draw,bin}=\min\left(N_{{\rm NL1-like,bin}},N_{{\rm BL1-like,bin}}\right).
\end{equation}
For each realization, we calculated the differences of the median spectral indices, $\Delta\alpha_\textrm{zmatch}$, as
\begin{equation}
\Delta\alpha_\mathrm{zmatch}={\rm median}(\alpha_{\rm NL1-like})-{\rm median}(\alpha_{\rm BL1-like})
\end{equation}
The more negative values of $\Delta\alpha_\mathrm{zmatch}$ indicate steeper median radio spectra for the NL1-like group. We repeated the procedure $5000$ times and characterized the resulting $\Delta\alpha_\mathrm{zmatch}$ distributions by their median and central 16--84 percentile interval. Unlike the fixed redshift cut, this procedure does not automatically discard all low-redshift sources. It retains the redshift bins in which both groups are represented and directly matches the numbers of objects from the two groups within each bin. It also quantifies the realization-to-realization variation of $\Delta\alpha_\mathrm{zmatch}$. 
To test the dependence of the results on the matching parameters, we repeated the analysis using redshift-bin widths of $\Delta z=0.10$, $0.15$, and~$0.20$,~requiring at least $1$, $3$, or~$5$ objects from each group in a retained bin. These tests determine whether the conclusions are sensitive to the adopted binning or to sparsely populated redshift bins. We found that the results did not depend on the choice of these parameters.

The redshift-matched $\Delta\alpha_\mathrm{zmatch}$ distributions obtained with the fiducial parameters are shown in Figure~\ref{fig:alpha_zmatched}, and~the corresponding numerical results are summarized in Table~\ref{tab:zmatched_nl_bl_delta}. The~redshift-matched median differences remain negative in all six comparisons, and~their central 16--84  percentile intervals lie entirely below zero. The~matched median differences are less negative than the observed values in all cases, indicating that differences between the redshift distributions may contribute to the measured effect but do not account for it completely. The~NL1-like samples retain systematically steeper radio spectra on average after their redshift distributions are directly matched to those of the corresponding BL1-like samples. The~ratios of the $\Delta\alpha_\mathrm{zmatch}$ to $\Delta\alpha_\mathrm{obs}$ range from $71$\% to $92$\%. The~ratios are closest to unity for the type 1 quasar subsamples ($\sim$90\%), while slightly lower for the Seyfert 1 subsamples ($\sim$80\%). Thus, the~spectral index difference for the Seyfert 1-like subsamples remains after redshift matching.

 \section{Sources with Large Spectral~Indices} \label{sec:alpha_outlier}

The two-point radio spectral index of a few objects exceeded the theoretical value of optically thick synchrotron emission, $2.5$. Below, we briefly present the radio properties of these objects and discuss the reasons behind their derived spectral index~values.

\subsection{NL1~AGN} \label{sec:outlierNLS1}

In the NL1 sample, SDSS\,J102818.15$+$535113.6 and SDSS\,J110546.07$+$145202.4 have spectral indices exceeding $2.5$. In~the case of SDSS\,J110546.07$+$145202.4, this is true for both of the VLASS observing epochs, while the $3$-GHz flux density of SDSS\,J102818.15$+$535113.6 decreased significantly between the two VLASS epochs. Its two-point radio spectral index is $1.4$ when the second-epoch VLASS data are~used.

SDSS\,J110546.07$+$145202.4 is a Seyfert 1 galaxy (located at a redshift of $z=0.12$) that showed an exceptional brightening by a factor of $>$20 at centimeter wavelengths during $\sim$18\,yr. Its multi-wavelength radio properties are discussed in detail in~\cite{J1105_2025}, and multiple radio, optical, UV, and~X-ray follow-ups and further discussion of outburst scenarios are presented in~\cite{Komossa2026,Dou2026}. 

SDSSJ\,102818.15$+$535113.6 has an absolute $B$-band optical magnitude of $-23.3$; thus, it formally can be regarded as a quasar. It is located at a redshift of $z=0.512$. Its flux density measured at $3$\,GHz with the VLASS changed significantly over the three epochs: $S^\mathrm{ep1}=(12.57 \pm 0.21)$\,mJy, $S^\mathrm{ep2}=(4.9 \pm 0.25)$\,mJy, $S^\mathrm{ep3}=(4.8 \pm 0.2)$\,mJy. It seems that the source underwent a brightening in $2017$ September during the first epoch of the VLASS observations; however, it faded back by 2020 August, when the second epoch of VLASS observations took place. In addition to~FIRST, it was also detected in the lower-resolution NRAO VLA Sky Survey, NVSS, \citep{nvss} with a flux density of $(2.5\pm0.4)$\,mJy. This slightly higher value than the one measured in FIRST, $(1.7 \pm 0.2 )$\,mJy, can be related to resolution issues or variability. We are not aware of more recent $1.4$-GHz flux density measurements of the object. At~lower frequencies, the~source was detected at $150$\,MHz with the LOFAR within the framework of the LoTSS DR2~\cite{lofar_dr2} with a flux density of $S^\mathrm{LOFAR}=(2.5 \pm 0.1)$\,mJy.

\subsection{BL1~AGN} \label{sec:outlierBLS1}

Here we present literature details on the eight BL1 AGN that have FIRST--VLASS spectral indices exceeding $2.5$. Half of these AGN showed such high spectral index values either using the first- or the second-epoch VLASS data. In~one case, SDSSJ111507.65$+$0237575, the~spectral index calculated with the second-epoch VLASS data is $2.3$. For~two other galaxies, SDSSJ094715.56$+$631716.4 and SDSSJ130414.70$+$591623.6, the~spectral indices calculated with the first-epoch VLASS data are $2.4$, and~$2.3$, respectively; thus, they formally do not exceed the limiting value of $2.5$.

The radio flux density values of the eight BL1 AGN measured in the FIRST, Rapid Australian Square Kilometre Array Pathfinder (ASKAP) Continuum Survey (RACS), and~VLASS are given in Table~\ref{tab:bls1_var}. The~first-epoch VLASS values are from the third pipeline version VLASS Quick Look catalog. For~the second and third epochs, we downloaded the available so-called `single epoch' (SE) images and~the quick-look images, respectively. The~SE images are reported to be of higher quality than the quick-look images taken in the same epoch~\cite{vlass_lacy}. For~SDSS\,J090007.99$+$364610.6, a~SE image was not available; therefore, we used the flux density value reported in the second pipeline version of the VLASS Quick Look catalog. To~obtain the flux densities, we fitted each image with single Gaussian brightness distribution models using the \textsc{imfit} task of the National Radio Astronomy Observatory (NRAO) Astronomical Image Processing System (\textsc{aips}) \cite{aips}. 

\begin{table}[H]
\small
\caption{Radio flux densities of BL1 AGN with $\alpha>2.5$. The~spectral index of the objects below the horizontal line exceed $2.5$ if the second-epoch VLASS observation is considered. Missing values indicate BL1 AGN not falling into the coverage of the given catalog. For~the second-epoch VLASS flux density, we give the value derived from the single-epoch~image.} \label{tab:bls1_var}
\begin{adjustwidth}{-\extralength}{0cm}
    \begin{tabularx}{\fulllength}{lccccccc}
    \toprule
    \textbf{Source Name} & \textbf{FIRST} & \multicolumn{3}{c}{\textbf{RACS}} & \multicolumn{3}{c}{\textbf{VLASS}} \\
    & \textbf{1.4\,GHz} & \textbf{0.889\,GHz} & \textbf{1.367\,GHz} & \textbf{1.655\,GHz} & \multicolumn{3}{c}{\textbf{3\,GHz}} \\
    & \textbf{1994--2009} & \textbf{2019--2020} & \textbf{2020--2022} & \textbf{2021--2022} & \textbf{Sep--Dec 2017} & \textbf{Aug--Oct 2020} & \textbf{Jan--Mar~2023}\\
    \midrule
     J001235.72$+$020326.3 & $0.8 \pm 0.1$ & $8.3 \pm 2.6$ & $3.5\pm 0.4$ & $5.1 \pm 0.7$ & $7.7\pm 0.3$ & $9.4 \pm 0.3$ & $12.7\pm 0.3$\\ 
     J001859.75$+$061931.9 & $7.0 \pm 0.2$ & $11.3 \pm 1.1$ & $29.9\pm1.8$ & $43.2\pm4.3$ & $74.3 \pm 0.3$ & $70.3\pm0.3$ & $77.4\pm0.3$ \\ 
           J023932.47$+$045805.9 & $0.5 \pm 0.1$ & --  & $2.0\pm 0.3$ & $2.5 \pm 0.4$ & $4.5\pm 0.2$ & $4.9\pm 0.3$ & $5.1 \pm0.3$\\ 
                  J090007.99$+$364610.6~$^{a}$ & $2.4\pm0.2$ & -- & $16.8\pm1.3$ & $26.7 \pm 2.7$ & $51.6 \pm 0.4$ & $63.6 \pm 0.2$ & $98.2\pm0.3$ \\ 
       J111507.65$+$023757.5~$^{b}$ & $1.6\pm 0.2$ & $1.96 \pm 0.5$ & $3.6\pm0.4$ & $6.4\pm 0.7$ & $12.5 \pm 0.2$ & $10.0\pm 0.3$ & $11.0\pm 0.2$\\ 
       J145958.43$+$333701.7 & $11.6 \pm 0.6$ & -- & $65.4\pm3.9$ & $75.7 \pm 8.1$ & $121.5\pm1.0$ & $90.7\pm0.2$ & $67.6 \pm 0.2$ \\ 
      \midrule
       J094715.56$+$631716.4 & $2.0 \pm 0.2$ & -- & -- & -- & $12.4 \pm 0.2$ & $20.2 \pm 0.2$ & $ 18.7 \pm 0.2$ \\
       J130414.70$+$591623.6 & $2.0 \pm 0.3$ & -- & -- & -- & $11.2 \pm 0.2$ & $15.6\pm 0.2~^{c}$ & $14.8\pm 0.2$ \\ 
       \bottomrule
    \end{tabularx}
    \end{adjustwidth}
    \noindent{\footnotesize{$^{a}$ The only galaxy for which a single-epoch VLASS image was not available; therefore, the flux density is obtained from the second-epoch VLASS quick-look catalog. Moreover, the epochs of the VLASS observations of this galaxy are different: May 2019, November 2021, and~July~2024. $^b$ The spectral index of this source is $2.3$ if the flux density of the second-epoch VLASS observation is~used.
    $^c$ The second-epoch VLASS observation took place on 27 June 2020.}}
\end{table}

The high spectral indices exceeding $2.5$ can be explained with the significant increase in the radio emission that happened between the FIRST and the VLASS observations or, alternatively, a~different emission mechanism, i.e.,~free-free absorption~\cite{freefree}. In~the case of six BL1 AGN (Table~\ref{tab:bls1_var}), the~comparison of FIRST and the more recent RACS~\cite{McConnell2020} data show the significant brightening at $\sim$1.4\,GHz; thus, flux density variability can explain the large spectral indices. Their radio brightenings range from a factor of $2$ to $7$. Except~for SDSS\,J023932.47$+$045805.9 and SDSS\,J111507.65$+$023757.5, the~sources also show significant variability at $3$\,GHz during the three VLASS~epochs. 

For two BL1 AGN, we are not aware of more recent $1.4$\,GHz radio flux density measurements; thus, there is no unambiguous evidence of flux density variability influencing the obtained spectral indices. Nevertheless, the~three epochs of VLASS data show the changes of the $3$-GHz flux density of these two sources as~well.

The brightest of these galaxies, SDSS\,J145958.43$+$333701.7, was studied in detail by~\cite{Orienti2020}. They reported its long-term flux density variability and the results of their VLBI observation of the source. The~mas-scale image can be interpreted as SDSS\,J145958.43$+$333701.7 being a compact symmetric object or a blazar. Owing to the large apparent speed measured for the radio features, Ref.~\cite{Orienti2020} concluded that the blazar hypothesis is more~likely.

 \section{\textbf{The Extremely Radio-Loud~AGN}} \label{sec:extremeRL}

Here, we comment on individual objects that stand out due to their radio loudness, specifically, with~$\log{\mathrm{(RL)}}>4$. In~particular, we also re-evaluate the optical NL1 classification of all systems based on literature data when available and~based on re-inspection of the optical SDSS spectra themselves. In~large samples, potential misclassifications can occasionally occur; in~particular, starburst galaxies or type 2 AGN can potentially be mistaken for NL1 active galaxies. Additionally, we checked VLBI archives to look for potential indications of relativistically beamed jet~emission. 

\subsection{NL1~AGN with $\log(\mathrm{RL})>4$}

\paragraph{SDSS\,J005833.80$+$062006.0 $(z=0.593)$: 
} The SDSS host galaxy image and the SDSS spectrum are very red. Broad H$\beta$ and \ion{Mg}{II} emission lines are present. Because~H$\alpha$ is redshifted out of the observable band, a~reliable extinction estimate is not possible. The~red broad-band spectrum clearly points to the presence of significant extinction. For~these reasons, we keep this galaxy as a good NL1 candidate in the sample, but~it is not included in the RL distribution investigation. We conclude that the radio loudness is overestimated, possibly by a high~factor.

Concerning its radio structure, it was detected with VLBI at $8$\,GHz as a single, mas-scale compact feature in several epochs with a median flux density of $\sim$170\,mJy~\cite{rfc_2025}.

\paragraph{SDSS\,J100888.04$+$073016.6:} This system is also known as 3C\,237. Its redshift ($z=0.068$) given in the SDSS DR16~\cite{sdssdr16},~used by~\cite{paliya2024}, is not the correct redshift. Likely, it arose from a misidentification of the redshifted [\ion{O}{II}]$3727$\,\AA\ emission line with H$\beta$. The correct redshift, which is reported in multiple data bases and~that properly matches the Balmer lines and all other emission lines, is $z=0.877$. In~fact, in the original catalog of 3C radio sources, the larger redshift, $z=0.877$ is given~\cite{3C_redshift}. This value is further used and confirmed by the spectral-line measurements of~\cite{Hirst_2003}. Additionally, the~SDSS DR13 online tool\endnote{\url{https://skyserver.sdss.org/dr13/en/tools/explore/summary.aspx} (accessed on 2 June 2026).} gives a very similar redshift value, $0.881$, reported within the framework of the Baryon Oscillation Spectroscopic Survey program~\cite{eBOSS_2016A}.

There is a detection of the H$\alpha$-[\ion{N}{II}] complex in the near-infrared spectrum~\cite{Hirst_2003}. The~whole unresolved complex has a $\mathrm{FWHM}=(2000\pm 500)$~km\,s$^{-1}$, still 
consistent with the idea that we see NLR emission. The~strong [\ion{O}{II}]$3727$\,\AA\, line and the red host galaxy suggest that we may see a low-ionization nuclear emission-line region (LINER) emission. The~Balmer lines appear as broad as forbidden lines, and~there is no \ion{Fe}{II} recognizable either. In~summary, many different criteria all falsify the NL1 classification of this object. Therefore, we removed it from the sample of NL1 AGN.

In the radio regime, 3C\,237 shows a $\sim$1.4$^{\prime}$-wide separated double-lobed feature at $1.6$\,GHz in the European VLBI Network (EVN) observation of~\cite{Fanti_237}. They concluded that the radio jets are most probably oriented close to the plane of the~sky. 

\paragraph{SDSS\,J140333.35$+$085944.0 $(z=0.45)$:} The SDSS host galaxy image and the SDSS spectrum are very red. Strong forbidden transitions are present, the~[\ion{O}{II}]$3727$\,\AA\, line is significantly stronger than the [\ion{O}{III}]$5007$\,\AA\, one, and~the Balmer lines are faint in comparison to the forbidden lines. This galaxy is most likely classified as a heavily reddened Seyfert type 1.9 with a strong starburst component. Because~of the heavy reddening, the~radio loudness of this system cannot be determined, and~we do not find positive evidence for its NLS1 classification. Therefore, we did not include this system in our analysis of NL1 galaxies.

Concerning its radio properties, it was observed with the Very Long Baseline Array (VLBA) at $5$\,GHz in 2015 as a single compact feature with $\sim$10\,mJy flux density. It is included in the Radio Fundamental Catalog~\cite{rfc_2025}. 

\paragraph{SDSS\,J161432.25$+$394444.8 $(z=0.891)$:} The broad-band SDSS spectrum of this galaxy is flat, missing any rise towards the blue. A~broad component in the H$\beta$ line is present, as~well as a highly broadened \ion{Mg}{II} emission line. Given the high redshift, the~spectrum is relatively noisy, and~a visual confirmation of the presence of \ion{Fe}{II} emission complexes is not possible. Because~H$\alpha$ is redshifted out of the observable band, a~reliable extinction estimate is not possible, even though the absence of a big blue bump is consistent with excess extinction. For~these reasons, we keep this galaxy as an NL1 candidate in the sample, but~it is not included in the RL distribution investigation. We caution that the radio loudness is likely strongly~overestimated.

This object was among the DEVOS targets observed with the EVN at $5$~GHz~\cite{devos2}. It was only marginally detected with $\sim$1~mJy flux density, at~a significant offset ($\sim$$0.7^{\prime\prime}$) from the SDSS optical position. The~EVN data indicate that the radio source is almost fully resolved on mas~scale.

\paragraph{SDSS\,J225221.75$-$004229.5 $(z=0.77)$:} This is another very red galaxy. There may 
be a faint broad Balmer component present in the spectrum and~broad \ion{Mg}{II} emission line, but~no \ion{Fe}{II} is recognizable. The~spectrum is more consistent with a heavily obscured intermediate-type Seyfert galaxy than an NLS1 galaxy. Therefore, we removed this system from our NL1 sample. Concerning its radio properties, it was detected as a single mas-scale compact feature of $40$\,mJy at $8$\,GHz in a VLBA observation in 2020~\cite{rfc_2025}.
  
  \subsection{BL1~AGN with $\log(\mathrm{RL})>4$}

\paragraph{SDSS\,J012156.85$+$042224.8 ($z=0.637$):} The SDSS spectrum shows a bona-fide type 1 AGN with strong broad and narrow emission lines. The~H$\beta$ profile indicates a type $1.5$ AGN. Because~the H$\alpha$ line is redshifted out of the observable band, a~reliable extinction estimate is not possible. Since the broad-band spectrum is slightly blue, we keep this galaxy as a BL1 in the sample. We caution that the radio loudness value might be~overestimated.

SDSS\,J012156.85$+$042224.8 is monitored at $15$\,GHz within the framework of the Monitoring of Jets of Active Galactic Nuclei with Very Long Baseline Array Experiments (MOJAVE) program~\cite{mojave_2018} and~shows a one-sided core--jet structure. It was also detected with the VLBI Space Observatory Program (VSOP) at $5$\,GHz, with~a core brightness temperature of $\sim$9 $\times 10^{11}$\,K~\cite{vsop}. Compared to the equipartition limit of intrinsic brightness temperature, $\sim$5 $\times 10^{10}$\,K~\cite{Readhead}, the~measured value implies relativistic beaming, the~object is a blazar, and its jet is seen at small angle to the line of~sight.

\paragraph{SDSS\,J012141.59$+$114950.4 ($z=0.57$):} The SDSS spectrum slightly rises towards the blue and~shows evidence for faint broad H$\beta$ emission, indicating a type 1.8 AGN and~strong broad \ion{Mg}{II} emission line. H$\alpha$ is just at the edge of the observable band so that an extinction estimate is not possible. This type 1 AGN spectrum may be reddened, and~we caution that the radio loudness may be~overestimated. 

SDSS\,J012141.59$+$114950.4 is also part of the MOJAVE sample~\cite{mojave_2018}. Its jet components show apparent superluminal motion with a speed of $\sim$19c (where $c$ denotes the speed of light), indicating a relativistically beamed jet emission observed at small angle to the line of sight~\cite{Lister_2019}. Thus, this source is also a~blazar.

\paragraph{SDSS\,J074125.73$+$270645.3 ($z=0.772$):} The SDSS image shows a blue host galaxy, and~the SDSS spectrum is blue, with~a strong broad H$\beta$ and \ion{Mg}{II} emission lines, implying a bona-fide type 1 AGN. Even though the H$\alpha$ line is redshifted out of the observable band and is not available for an extinction estimate, this AGN does not appear to be heavily reddened and the radio loudness estimate is therefore~reliable.

According to $5$- and $8$-GHz VLBI observations of~\cite{Tremblay_2016}, this object shows a flat-spectrum core and two-sided, steep-spectrum jet structure at mas-scale resolution. It is classified as a compact symmetric object~\cite{GPS_review}; thus, its radio emission is not dominated by relativistic~beaming.

\paragraph{SDSS\,J080413.87$+$470442.8 ($z=0.51$):} This object is also known as 4C\,47.27. The~SDSS image shows a red host galaxy, and~the SDSS spectrum is flat. It is dominated by strong [\ion{O}{III}] line emission; a~faint broad component of H$\beta$ is recognizable, as well as strong broad \ion{Mg}{II} emission. The~H$\alpha$ line is unobservable. This type 1 AGN spectrum may be reddened, and~we caution that the radio loudness may be~overestimated. 

According to the $8$-GHz VLBI observation of~\cite{Dallacasa_2002}, the~mas-scale radio emission originates from a region of $\sim$$1^{\prime\prime}$ containing a bent jet and a faint~core.

\begin{adjustwidth}{-\extralength}{0cm}

\printendnotes[custom] 

\reftitle{References}



\PublishersNote{}
\end{adjustwidth}
\end{document}